\documentclass[epj]{svjour}
\usepackage{graphicx}
\usepackage{epsfig}
\begin{document}
\title{The activation method for cross section measurements in nuclear astrophysics}
\author{Gy. Gy\"urky\inst{1} \and Zs. F\"ul\"op\inst{1} \and F. K\"appeler\inst{2} \and G.G. Kiss\inst{1} \and A. Wallner\inst{3} 
}                     
%
%
\institute{Institute for Nuclear Research (Atomki), P.O.B 51, H-4001 Debrecen, Hungary \and  Karlsruhe Institute of Technology, Hermann-von-Helmholtz-Platz 1 76344 Eggenstein-Leopoldshafen, Germany \and Research School of Physics and Engineering, The Australian National University, ACT 2601, Australia}
\date{Received: date / Revised version: date}
%
\abstract{
The primary aim of experimental nuclear astrophysics is to determine the rates of nuclear reactions taking place in stars in various astrophysical conditions. These reaction rates are important ingredient for understanding the elemental abundance distribution in our solar system and the galaxy. The reaction rates are determined from the cross sections which need to be measured at energies as close to the astrophysically relevant ones as possible. In many cases the final nucleus of an astrophysically important reaction is radioactive which allows the cross section to be determined based on the off-line measurement of the number of produced isotopes. In general, this technique is referred to as the activation method, which often has substantial advantages over in-beam particle- or $\gamma$-detection measurements. In this paper the activation method is reviewed from the viewpoint of nuclear astrophysics. Important aspects of the activation method are given through several reaction studies for charged particle, neutron and $\gamma$-induced reactions. Various techniques for the measurement of the produced activity are detailed. As a special case of activation, the technique of Accelerator Mass Spectrometry in cross section measurements is also reviewed.
\PACS{
      {26.20.-f}{Hydrostatic stellar nucleosynthesis}   \and
      {26.30.-k}{Nucleosynthesis in novae, supernovae, and other explosive environments} \and
			{24.10.-i}{Nuclear reaction models and methods} 
     } 
} 
\maketitle
\setcounter{tocdepth}{4}
\tableofcontents

\section{Introduction}
\label{sec:intro}

In 2017 the 60$^{\rm th}$ anniversary of two seminal publications was celebrated which is considered to be the birth of a new scientific discipline called nuclear astrophysics. In 1957 E.M. Burbidge, G.R. Burbidge, W.A. Fowler and F. Hoyle (often abbreviated as B$^2$FH) \cite{RevModPhys.29.547} and independently A.G.W. Cameron \cite{CameronCRR} gave a detailed description of nuclear processes in stars and explained many aspects of astronomical observations and the abundances of chemical elements and their isotopes observed in nature.

Our understanding of nuclear processes in our Universe has been improved tremendously since the publication of these papers. This is in part due to the fact that many nuclear reactions of astrophysical importance have been studied experimentally providing input for the astrophysical models. The activation method has played a vital role in these experiments. This method is reviewed in the present paper with regard to nuclear astrophysics experiments. First a short introduction to nuclear astrophysics and to the activation method is given.

\subsection{A short account of nuclear astrophysics and the importance of cross section measurements}

It is a common knowledge that the only source of energy that can power stars for billions of years is nuclear energy \cite{celnikier2006find}. The fusion of hydrogen into helium generates enough energy for billions of years in main sequence stars like our Sun. After the exhaustion of hydrogen, a gravitational contraction will occur until He-burning starts to be efficient in the solar core. Thus, production of carbon and oxygen will be ignited before the Sun runs out of fuel. More massive stars can go through a series of burning stages that allow them to sustain their operation, however, for shorter and shorter periods of time, a sequence that eventually may lead to a supernova explosion \cite{2002RvMP74.1015W}. All the quiescent burning stages and also the stellar explosions involve nuclear reactions. Information about these reactions is of fundamental importance for the detailed understanding of these processes. 

Nuclear reactions are not only responsible for the energy generation of stars, but also for the synthesis of the chemical elements that are observed in the universe today. Primordial nucleosynthesis, one of the strongest pieces of evidence for the Big Bang theory, relies on the knowledge of about a dozen of reactions that took place in the first few minutes after the Big Bang \cite{RevModPhys.88.015004}. Hydrogen, Helium and a small amount of Lithium were produced in this way, while heavier elements could only be made during stellar evolution. Li, Be and B represent an exception since, besides lithium production during Big Bang nucleosynthesis, their primary production source is galactic cosmic-ray nucleosynthesis \cite{RevModPhys.66.193}.

The stellar burning phases of stars build chemical elements up to the Iron group by sequences of charged particle induced reactions. The chemical elements heavier than Iron are synthesized in stars in processes taking place in parallel with a given main stellar burning process. The dominant fraction of the heavy elements is thought to be produced by sequences of neutron capture reactions. Two main distinct processes are considered, the s-process which takes place in giant stars \cite{RevModPhys.83.157} and the r-process which can occur only in explosive stellar environments \cite{Arnould200797} and/or in neutron star mergers \cite{1538-4357-525-2-L121}. Other neutron capture processes like the i-process are also suggested to explain part of the observed heavy element abundances \cite{0004-637X-831-2-171}. The production mechanism of the heavy, proton-rich isotopes that cannot be synthesized by neutron-induced reactions, is in general referred to as the astrophysical p-process \cite{2003PhR3841A}. Various sub-processes are considered, the most important role is attributed to the $\gamma$-process \cite{0034-4885-76-6-066201}. 

Nuclear reactions play a key role in all processes of energy generation and nucleosynthesis. With the exception of cosmic-ray induced reactions, the reactions take place in a plasma environment at thermal equilibrium. The interaction energies are therefore determined by the temperature of the plasma. The key quantity that determines the energy release from a given reaction and the rate of production or destruction of a given isotope is the thermonuclear reaction rate. In the case of two reacting particles (none of them is a $\gamma$-quantum) the rate is given by the following formula \cite{rolfs1988cauldrons,iliadis2015nuclear}:

\begin{equation}
	r_{1,2}=N_1N_2 \int_0^\infty{v\sigma(E)P(E)dE}
\end{equation}
where $N_1$ and $N_1$ are the number densities of the two reacting particles in the plasma, $v$ and $E$ are, respectively, the relative velocity and energy of the particles, $P(E)$ is the energy distribution and $\sigma(E)$ is the cross section of the reaction as a function of the interaction energy. At typical stellar conditions the energy distribution of the interacting particles can be well approximated by a Maxwell-Boltzmann distribution. In this case the reaction rate formula becomes

\begin{equation}
	r_{1,2}=N_1N_2 \left( \frac{8}{\pi \mu} \right)^{1/2}\frac{1}{kT^{3/2}}\int_0^\infty{E\sigma(E)e^{-E/kT}dE}.
\label{eq:rate2}
\end{equation}
Here T is the plasma temperature and $\mu$ is the reduced mass of the reacting particles. 

In order to calculate the reaction rate, information about the cross section as a function of energy is needed. In fact, the cross section must be known only within a limited energy range where the integrand in Eq.\,\ref{eq:rate2} is not negligible. In the case of charged particle induced reactions the important energy region is called the Gamow-window \cite{iliadis2015nuclear,PhysRevC.81.045807} which results from the combination of the Maxwell-Boltzmann distribution and the energy dependence of the cross section. Owing to the Coulomb barrier penetration effect, the Gamow-window is shifted to much higher energies than the thermal energy kT. Two examples: for the Hydrogen burning reaction $^3$He($\alpha,\gamma$)$^7$Be at solar temperature of 15\,MK (kT\,=\,1.3\,keV) the Gamow-window is between about 15 and 30\,keV \cite{Bordeanu20131}. At 2\,GK temperature (kT\,=\,170\,keV) encountered in a supernova explosion, the Gamow-window for the $^{130}$Ba($\alpha,\gamma$)$^{134}$Ce reaction relevant to the $\gamma$-process is between about 5 and 8\,MeV \cite{PhysRevC.85.025804}. On the contrary, the relevant energy for neutron-induced reactions lies very close to the thermal kT one, because no Coulomb repulsion is present between the interacting particles \cite{0954-3899-41-5-053101}.

Knowledge collected over several decades about astrophysical processes and the physics of stars would not have been possible without the knowledge about stellar reaction rates and therefore about cross sections \cite{Bertulani201656}. In principle, nuclear theory can provide the necessary cross sections, but for increasing the reliability of the models, accurate experimental cross section data are indispensable. In many cases, especially for charged particle induced reactions, the extremely low (sub-$\mu$barn) cross section at astrophysical energies prevents direct measurements. However, measurements at higher energies help constrain nuclear theory and more reliable reaction rates can be expected if experimental information backs up theory. As nuclear astrophysics is a quickly evolving field, new and often more precise experimental data on astrophysically important reactions are continuously needed \cite{0034-4885-74-9-096901}. Therefore, as they were in the past, cross section measurements will also be a hot topic in the future.

It is worth noting that for the calculation of the reaction rate, the total reaction cross section is needed. Angle-differential cross sections are not of direct relevance for astrophysics. Similarly, the way the final nucleus is created, i.e. the pattern of various transitions leading to its ground state, plays no role. One experimental technique which directly provides the total, angle integrated production cross section is the activation method. The general features of this method, which has other advantages from the viewpoint of nuclear astrophysics, will be reviewed in the next section.

\subsection{The activation method for cross section measurements}
\label{sec:actmethod}

In a typical cross section measurement of a nuclear reaction $A(b,c)D$, a target containing nuclei $A$ with known and homogenous surface density $N_A$ (atoms/cm$^2$) is bombarded by a known current $\Phi_b$ (1/s) of beam particles $b$ (which can also be photons, see sect.\,\ref{sec:gammainduced}). It is assumed that the lateral size of the beam is smaller that the target area. The cross section $\sigma$ (cm$^2$) is then determined by measuring the number of reactions $N_{reac}$ (1/s) that take place \cite{Knoll:1300754},

\begin{equation}
	\sigma_{reac} = \frac{N_{reac}}{N_A \cdot \Phi_b}
\label{eq:crosssec}
\end{equation}

There are different possibilities for the determination of $N_{reac}$. Perhaps the most common method in nuclear physics and also in nuclear astrophysics is the detection of the light outgoing particle $c$ of the reaction. Since nuclear reactions typically occur on very short time scales (10$^{-15}$--10$^{-22}$\,s), this technique can be referred to as the prompt or in-beam method as the detection of the outgoing particles must be carried out during the beam bombardment. The in-beam method is often a valuable complementary approach to the activation described below as it allow to measure partial cross sections and to examine reactions where the reaction product is not suitable for activation experiments.

The other possibility is the determination of the number of produced heavy residual nuclei $D$. Often these heavier reaction products $D$ carry a relatively small kinetic energy compared to the beam particles $b$ and may not even leave the target. Their prompt detection requires special experimental techniques such as a recoil separator combined with an inverse kinematics experiment (i.e. the light particle $b$ is bombarded by the heavy nucleus $A$) \cite{Gross200012,Hutcheon2003190,SCHURMANN2004428,Couder200835,BERG2016165}.  

If the produced heavy residual nuclei $D$ are radioactive, their number can be determined via their decay. This is the basis of the activation method. Let us suppose that the radioactive species created has a decay constant $\lambda$ [1/s] which is related to its half-life: $t_{1/2}$\,=\,ln(2)/$\lambda$ [s]. If the target is irradiated by a beam with constant current for a time period of $t_{irrad}$, then the number of produced radioactive nuclei still alive at the end of the irradiation is given by:
\begin{equation}
	N_{prod} = \sigma_{reac} \cdot N_A \cdot \Phi_b \cdot \frac{1-e^{-\lambda t_{irrad}}}{\lambda}
\label{eq:production1}
\end{equation}
where the last exponential term accounts for the decay of the reaction product during the irradiation. It converges to the time duration of the irradiation $t_{irrad}$ if the half-life of the reaction product is much longer than $t_{irrad}$. If the half-life is shorter than or comparable to $t_{irrad}$ and the condition of constant beam current during the irradiation is not fulfilled, then the irradiation period must be divided into $n$ sufficiently short intervals during which the current can be regarded as constant. Then the above formula becomes:
\begin{equation}
	N_{prod} = \sigma_{reac} \cdot N_A \cdot \sum_{i=1}^n \Phi_{b,i} \cdot \frac{1-e^{-\lambda \tau}}{\lambda} \cdot e^{-\lambda \tau (n-i)}
\label{eq:production2}
\end{equation}
where $\tau\equiv t_{irrad}/n$ is the length of the time period and $\Phi_{b,i}$ is the beam current in the $i^{\rm th}$ period. The last exponential term takes into account the decay of the produced isotopes between the $i^{\rm th}$ period and the end of the whole irradiation. 

\begin{figure}
	\includegraphics[angle=270,width=\columnwidth]{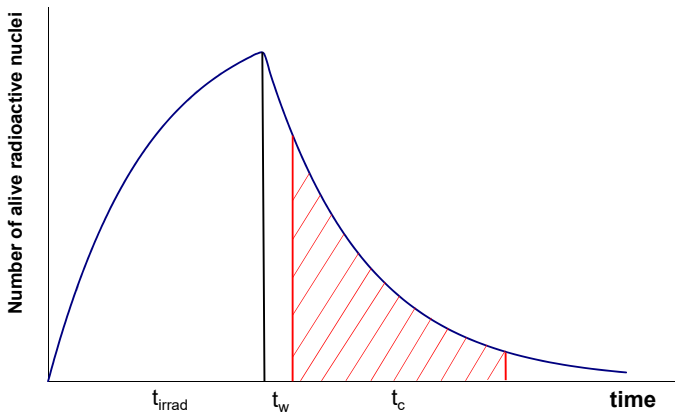}%
	\caption{Schematic illustration of an activation process. The number of alive radioactive nuclei is plotted as a function of time. In this example the target is irradiated for three half-lives and the decay is also measured for three half-lives after a 0.5 half-life waiting time.}
	\label{fig:production_decay}
\end{figure}

For the cross section measurement $N_{reac}$ must be determined. This can be achieved by observing the decay of the reaction product for a given counting time $t_{c}$. The number of decays is given by the following formula \cite{Knoll:1300754}:
\begin{equation}
	N_{decay} = N_{prod} \cdot e^{-\lambda t_w} \cdot (1-e^{-\lambda t_c})
\label{eq:decay}
\end{equation}
where $t_w$ is the waiting time elapsed between the end of the irradiation and the beginning of the counting. The production and decay of the reaction product is illustrated schematically in fig.\,\ref{fig:production_decay}.

If the half-life is much longer than any reasonable counting time $t_c$, then $N_{decay}$ will be very small making the cross section determination difficult or impossible. In such a case, a different determination of $N_{prod}$ is necessary. The method of Accelerator Mass Spectrometry (AMS) can be applied in the case of long half-lived reaction products. This technique will be reviewed in sect.\,\ref{sec:AMS}

An activation experiment can be divided into two separate phases: the production of the radioactive reaction products, i.e. the irradiation phase, and the measurement of the decay of the reaction products. In the next sections, the experimental techniques used in an activation measurement in nuclear astrophysics will be discussed in detail separately for the two phases:
\begin{itemize}
\item Section \ref{sec:firstphase} deals with the irradiation (production) phase.
\item Section \ref{sec:secondphase} deals with activity measurements.
\item Section \ref{sec:AMS} describes the direct atom counting of the reaction products.
\end{itemize}

Several examples will be used to discuss the various experimental aspects. Related to different astrophysical processes, these examples will be taken mainly from experiments carried out with the participation of the authors of the present review. In stellar hydrogen burning there are a few reactions which lead to radioactive isotopes and the activation method can therefore be used. For example, one of the key reactions of the pp-chain of hydrogen burning is $^3$He($\alpha,\gamma$)$^7$Be which leads to the radioactive $^7$Be. This reaction was studied many times using the activation method \cite{Bordeanu20131,PhysRevLett.48.1664,PhysRevC.27.11,1983ZPhyA.310...91V,PhysRevLett.93.262503,PhysRevLett.97.122502,PhysRevC.75.035805,PhysRevC.76.055801,PhysRevC.86.032801}. Other examples in hydrogen burning are the $^{17}$O(p,$\gamma$)$^{18}$F \cite{PhysRevLett.95.031101,PhysRevC.75.035810,PhysRevLett.109.202501,PhysRevC.89.015803,PhysRevC.95.035805} or $^{14}$N(p,$\gamma$)$^{15}$O \cite{doi:10.7566/JPSCP.14.020403} reactions. Other astrophysical process for which the activation method is extensively used is the $\gamma$-process \cite{0034-4885-76-6-066201}. The dominant part of the experimental cross section database \cite{SZUCS2014191} relevant to the $\gamma$-process is collected from activation experiments \cite{PhysRevC.85.025804,PhysRevC.55.3127,PhysRevC.58.524,1998A&A.333.1112S,1999NuPhA.652..391C,PhysRevC.64.065803,2002NuPhA.710..469O,PhysRevC.66.015803,PhysRevC.68.055803,PhysRevC.74.025805,0954-3899-34-5-003,PhysRevC.76.055807,PhysRevC.75.025801,PhysRevC.75.015802,PhysRevLett.101.191101,PhysRevC.78.025804,a4f5e5a33acd43bfa952482c6f955508,PhysRevC.79.065801,PhysRevC.80.035804,0954-3899-37-11-115201,Kiss2011419,PhysRevC.84.015802,PhysRevC.83.064609,PhysRevC.84.045808,PhysRevC.86.041601,PhysRevC.86.035801,PhysRevC.85.028801,Netterdon2013149,PhysRevC.90.035806,PhysRevC.90.052801,Gyurky2014112,Kiss201440,PhysRevC.90.065807,PhysRevC.91.055809,0954-3899-42-5-055103,PhysRevC.91.034610,PhysRevC.94.045801,PhysRevC.93.025801,SCHOLZ2016247,PhysRevC.96.045805}. In the case of neutron induced reactions, the ones involved in the s-process network are often studied by the activation method \cite{RevModPhys.83.157}. 

Before a detailed description of the nuclear astrophysics motivated activation experiments, it is worth noting that activation is a widely used technique in many applications, e.g. material analysis. The activation analysis determines the elemental and isotopic composition of unknown samples based on the known cross section of a given reaction and on the detection of the decay of the reaction products. This is in contrast with the activation cross section measurements where the cross section is the unknown quantity which is to be determined using a sample of known composition. Neutron activation analysis is a particularly powerful technique as neutrons can penetrate deep into the samples and provide therefore information about the bulk and not only about the surface. Charged particle activation analysis, on the other hand, is important in cases where only the surface layers of the samples are to be studied. Exhaustive information can be found in many available textbooks about the activation analysis method \cite[and many others]{vertes2010handbook,Kugler,Verma2007,Molnar}. 

\section{First phase of an activation experiment: production of the radioactive species}
\label{sec:firstphase}

In the first phase of an activation cross section measurement, a suitable target is bombarded by a beam of projectiles. The target properties (thickness, composition, etc.) must be known. The determination of the cross section requires the knowledge of the number of target nuclei (an important exemption is the AMS method, see section \ref{sec:AMS}) and the number of projectiles impinging on the target (see eq.\,\ref{eq:crosssec}). In nuclear astrophysics, charged particle (mostly proton or alpha), neutron and $\gamma$-induced reactions are studied. The activation by theses three types of beams requires different experimental conditions. These special conditions are discussed separately in the following subsections. 

\subsection{Charged particle induced reactions}
\label{sec:chargedinduced}

\subsubsection{Relevant beam energies}
As emphasized in the introduction, proton and alpha-induced reactions play a key role in many astrophysical processes from hydrogen burning up to the processes taking place in e.g. a supernova explosion. The cross section of these reactions must be known at relatively low energies corresponding to the Gamow-window or as close to it as possible. Additionally, a wide energy range is often mandatory for the cross section measurements in order to enable reliable theory-based extrapolation to the energies of astrophysical relevance. 

The number of energy points measured within the chosen energy range, i.e. the resolution of the excitation function depends on the expected energy dependence of the cross section. Where a smooth variation of the cross section is expected as a function of energy, fewer points will be enough to constrain theoretical cross section calculations \cite[e.g.]{PhysRevC.94.045801}. Where the cross section is dominated by narrow resonances, fine energy steps and/or resonance strength determinations are necessary \cite[e.g.]{PhysRevC.95.035805}. In some cases, the assumption of smoothly varying cross section proves to be wrong and stronger fluctuations in the excitation function are observed. This means that more energy points of the cross sections have to be measured in order to provide reliable reaction rates. A good example is the $^{92}$Mo(p,$\gamma$)$^{93}$Tc reaction relevant to the $\gamma$-process where the low level density of the neutron-magic reaction product means that the basic assumption of the statistical model is not valid. The observed fluctuation and the disagreement between the available experimental data indicate the need for further studying this reaction. See ref. \cite{Gyurky2014112} and references therein.

\subsubsection{Irradiations}

Low-energy proton or alpha beams are typically provided by electrostatic accelerators or cyclotrons. In order to measure low cross sections, high beam intensities are usually required to produce sufficient reaction products (see, however, sect. \ref{sec:chargedtarget} for target stability issues). Following eq.\,\ref{eq:crosssec}, for the cross section determination the number of projectiles impinging on the target must be known. In the case of charged particle induced reactions, this number can easily be obtained based on charge measurement (with the exception of gas targets, see sec.\,\ref{sec:gastarget}). The chamber where the target is placed must form a good Faraday cup so that the measurement of electric current delivered by the beam to the target can be converted into the number of projectiles. An example of a typical activation chamber can be seen e.g. in fig. 1. of ref.\,\cite{PhysRevC.91.034610}.

The duration of the irradiations is typically defined by the half-life of the reaction product studied. As the number of produced isotopes goes to saturation (see eq.\,\ref{eq:production1}), irradiations longer than about three half-lives do not provide any additional yield. Often more than one reaction, i.e. more than one reaction product is studied in one experiment. In such a case the longest half-life determines the length of the irradiation. If the half-lives are long, in the range of days, or longer, then the duration of the activation is usually limited by the available accelerator time.

If the length of the irradiation is longer than or comparable to the half-life of a reaction product then the variation of the beam intensity must be recorded and taken into account in the analysis. This is usually done by recording the charge on the target as a function of time using the multichannel scaling (MCS) mode of an ADC. The time basis of the MCS is determined by the respective half-lives. 

\begin{figure}
	\includegraphics[angle=270,width=\columnwidth]{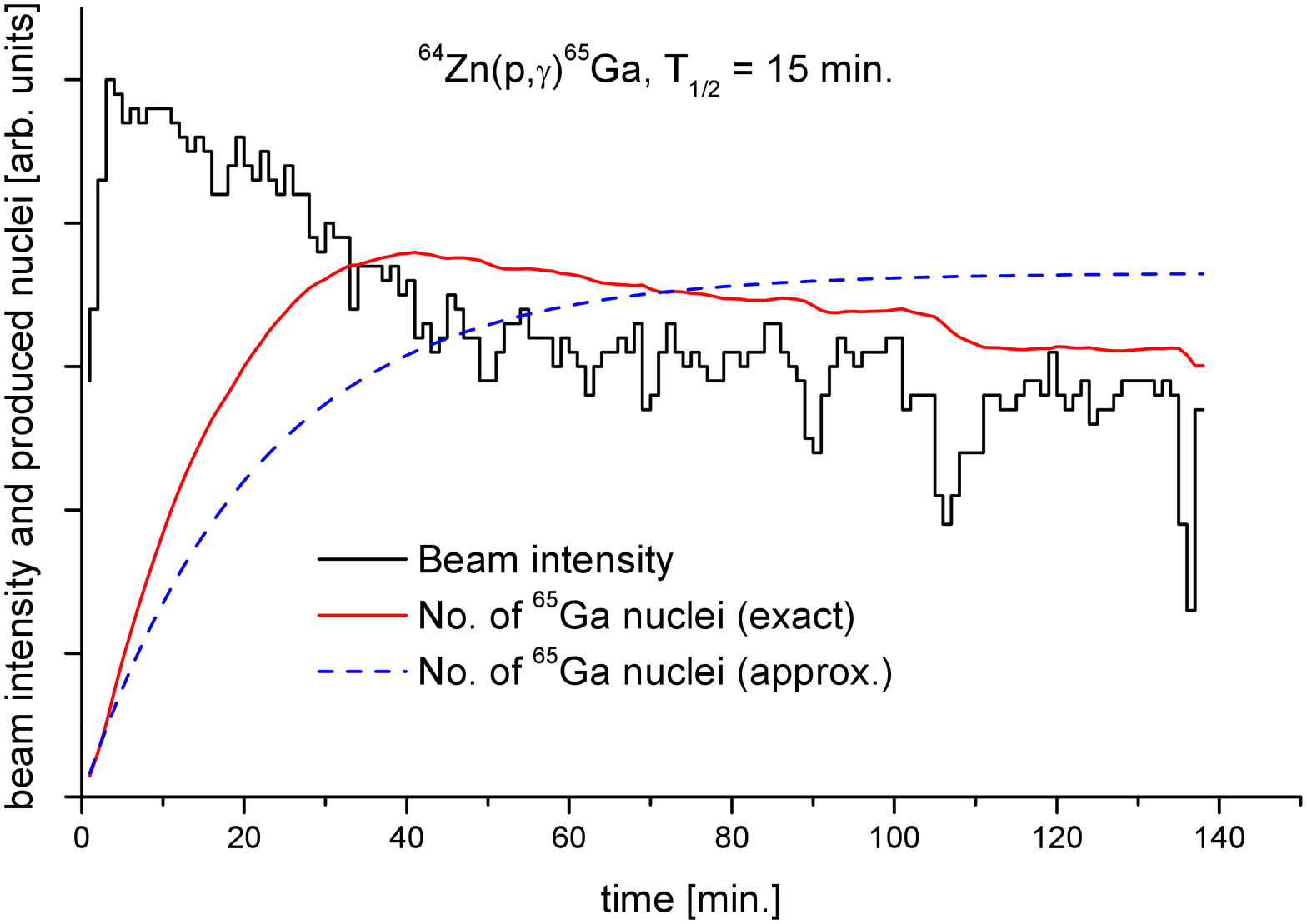}%
	\caption{Variation of the beam intensity with 1 minute time intervals. The number of alive $^{65}$Ga nuclei produced in the $^{64}$Zn(p,$\gamma$)$^{65}$Ga reaction is also shown calculated based on the measured beam intensity variation (exact) and supposing constant beam intensity (approximated). It is clearly seen that the assumption of the constant beam intensity leads to a wrong estimation of the produced isotopes. The y axis is in arbitrary units.}
	\label{fig:beam_intensity_variation}
\end{figure}

The necessity of MCS charge recording is illustrated in fig.\,\ref{fig:beam_intensity_variation}. The shown histogram of the proton beam intensity was recorded during the measurement of the $^{64}$Zn(p,$\gamma$)$^{65}$Ga cross section \cite{PhysRevC.90.052801}. It can be seen that the beam current was fluctuating and decreasing continuously during the activation time of 140 minutes. The two curves in the figure show the calculated number of live $^{65}$Ga nuclei (in arbitrary units) based on the recorded MCS histogram and based on the assumption of constant (averaged) beam current during the activation. Not recording the beam current variation would result in an overestimation of reaction products by about 20\,\% in this case as it can be seen from the different values of the red and blue curves at the end of the irradiation. It should be noted that the long irradiation was necessary as in this experiment besides the $^{64}$Zn(p,$\gamma$)$^{65}$Ga reaction, the $^{64}$Zn(p,$\alpha$)$^{61}$Cu reaction channel was also measured, where the half-life of $^{61}$Cu is much longer, 3.4 hours compared to the 15.2\,min half-life of $^{65}$Ga.

\subsubsection{Target properties and characterizations}
\label{sec:chargedtarget}

Charged particles lose energy quickly when passing through matter. In order to obtain the cross section at a well defined energy, thin targets must be used where the energy loss of the beam is small compared to the characteristic variation of the cross section as a function of energy. This requirement is fulfilled with targets having a thickness up to a few times 10$^{18}$ atoms/cm$^2$, which corresponds to a thickness of typically of order of 100\,nm. Such thin layers are normally produced by vacuum evaporation or sputtering techniques onto a support material (backing) \cite{evaporation}. In some cases, special techniques are needed such as anodization for producing oxygen targets, for example for the $^{17}$O(p,$\gamma$)$^{18}$F reaction \cite{Caciolli2012}.

For an activation experiment the backing of the target can be either thin or thick type. By definition, a thick backing completely stops the beam. A thin backing, on the other hand, allows the beam to pass through losing only a small fraction of its energy. The beam stop, where the charged particle beam is fully stopped, can be independent in this case having some advantages for reducing beam induced background (see sect.\,\ref{sec:background}). A thin target backing must be thick enough to fully stop the radioactive reaction products, as the induced activity is measured later in the target itself. In typical cases, however, based on the reaction kinematics the reaction products have such low energies that they are fully stopped in the target backing foils which normally have thicknesses of the order of micrometers. 

In addition to the number of projectiles, the number of target atoms, more correctly the surface density of target atoms, must also be known. This quantity is referred to as the target thickness. There are different ways to determine the target thickness. Perhaps the easiest way is to directly measure the weight of the target backing before and after the deposition of the target layer. This method can be used only for thin backings. Furthermore, the stoichiometry of the target material must be known a priori as weighing does not give any information about the molecular composition. This method is best used in the case of single element targets. 

Other methods for the determination of target thickness usually involve ion beam analysis techniques \cite{Verma2007b}. The most commonly used methods are Rutherford Backscattering Spectroscopy (RBS), Particle Induced X-ray Emission (PIXE) and Nuclear Reaction Analysis (NRA). These measurements for the target thickness determination are usually carried out before the actual activation cross section measurement. It has been shown that an inspection of the target before and after the activation process can also be useful (see below). In the next paragraphs one example will be given for each of these methods. 

\begin{figure}
	\includegraphics[angle=270,width=\columnwidth]{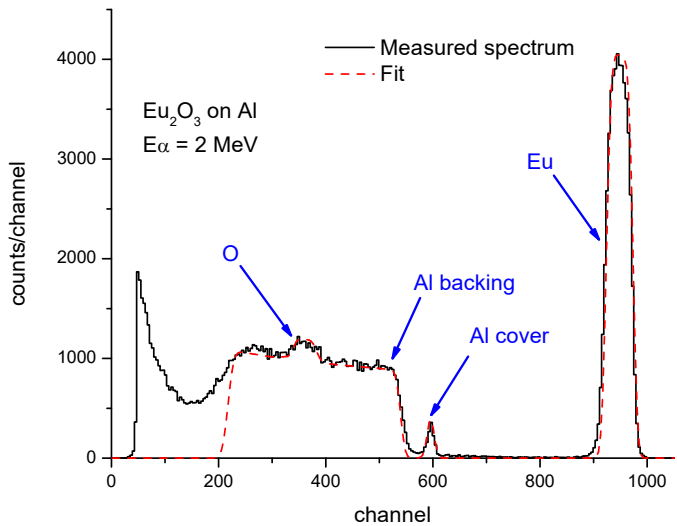}%
	\caption{RBS spectrum of a Eu$_2$O$_3$ target evaporated onto a 2\,$\mu$m thick Al foil.}
	\label{fig:Eu_RBS}
\end{figure}

RBS is a very powerful method for determining the absolute target thickness if the chemical element to be studied is well separated in the spectrum from the other elements in the target or backing. The best result will be obtained for a heavy element target that is deposited on a backing made of light elements. Figure\,\ref{fig:Eu_RBS} shows an RBS spectrum of a Eu$_2$O$_3$ target evaporated onto a 2\,$\mu$m thick Al foil \cite{0954-3899-37-11-115201}. For the thickness measurement, a 2\,MeV $\alpha$-beam was used and the backscattered alpha particles were detected at an angle of 165$^\circ$ with respect to the beam direction. The $\alpha$-particles scattered by the heavy Eu can be easily distinguished from those scattered on Al or O. The spectrum was fitted using the SIMNRA code \cite{SIMNRA}. As the area of the Eu peak can be related to the height of the plateau of the Al backing, the absolute Eu thickness can be determined independently from the number of $\alpha$-particles hitting the target and from the solid angle covered by the particle detector \cite{PhysRevC.55.3127}. Systematic uncertainties related to the charge integration and solid angle determination are thus avoided. Therefore, more precise target thickness measurements can be carried out and relatively simple experimental setups can be used. 

\begin{figure}
	\includegraphics[angle=270,width=\columnwidth]{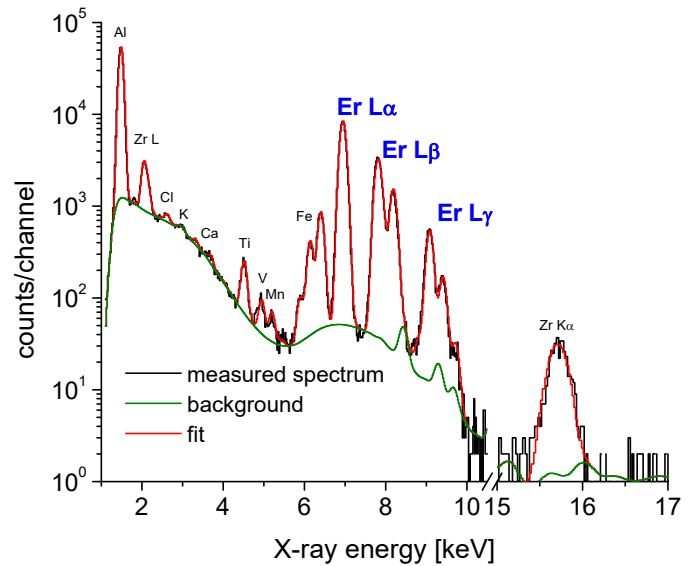}%
	\caption{PIXE spectrum of an Er target evaporated onto a 2\,$\mu$m thick Al foil.}
	\label{fig:PIXE_162Er}
\end{figure}

The PIXE technique is very sensitive and allows the determination of trace element concentrations in various samples \cite{PIXE}. It can, however, also be used for the quantitative determination of the amount of chemical elements making up the target in a sample (i.e. for a target thickness measurement) if the layer is suitably thin so that the X-ray self-absorption can be controlled. Figure\,\ref{fig:PIXE_162Er} shows a typical PIXE spectrum of an Er target evaporated onto a thin Al foil. Besides the X-ray peaks corresponding to Al and Er, other peaks originating from trace element impurities in the sample are also labeled. This indicates the high sensitivity of the method. For an absolute determination of the target thickness the number of projectiles hitting the target and the absolute efficiency of the X-ray detector must be known. Therefore, the target thickness measurement with the PIXE technique usually requires a dedicated PIXE setup \cite{BORBELYKISS1985496}.

\begin{figure}
	\includegraphics[angle=270,width=\columnwidth]{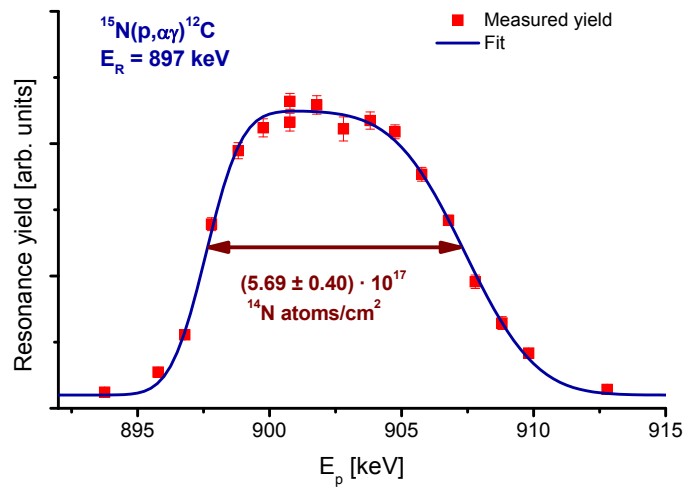}%
	\caption{Resonance profile measured on a TiN target deposited onto a thick Ta backing using the E$_p$\,=\,897\,keV resonance in $^{15}$N(p,$\alpha\gamma$)$^{12}$C reaction.}
	\label{fig:N_NRA}
\end{figure}

If a suitable nuclear reaction can be induced on the target isotope to be studied, the NRA method can be used for target thickness determination \cite{NRA}. This method is especially powerful if a narrow resonance is present in the studied reaction (resonant NRA). In this case only the resonance profile of the target must be measured, i.e. the yield of the resonance as a function of the bombarding energy around the resonance. There is no need for the precise knowledge of the reaction cross section or the resonance strength. Only the composition of the target (the stoichiometry) and the stopping power must be known. 

An example is shown in fig.\,\ref{fig:N_NRA}. A TiN target sputtered onto a thick Ta backing \cite{doi:10.7566/JPSCP.14.020403} was investigated using the E$_p$\,=\,897\,keV resonance in the $^{15}$N(p,$\alpha\gamma$)$^{12}$C reaction \cite{PhysRevC.81.055807}. Having independent information about the Ti:N ratio in the target and knowing the proton stopping power in Ti and N, the target thickness can be obtained from the width of the measured resonance profile.

Besides these three methods, there are several other techniques which can be used for target thickness determination. These include, but are not limited to, X-Ray Fluorescence Analysis (XRF) \cite{beckhoff2007handbook}, Elastic Recoil Detection Analysis (ERDA) \cite{ERDA} and Secondary Ion/Neutral Mass Spectrometry (SIMS/SNMS) \cite{Vad200913}. Often it is useful to carry out two or more independent measurements of the target thickness in order to increase the reliability of this important quantity. This is especially necessary with targets having an uncertain stoichiometry \cite{Ba_target}. The final uncertainty of the measured target thickness will strongly depend on the properties of the target itself and on the methods used. Using more than one method is useful also for reducing the uncertainty.

Since low cross section measurements require high beam intensities (often hundreds of microamperes), the possibility of target degradation under beam bombardment must be studied. As opposed to the in-beam experiments, in activation there is no continuous measurement of the reaction yield during the irradiation, any target degradation must thus be checked independently. One possibility is to measure the target thickness both before and after the irradiations and if no difference is found, the target stability is guaranteed. In such cases the thickness measurement must be carried out precisely on the beam spot irradiated during the activation. With this method, however, the degradation of the target is revealed only after the irradiation and - in the case of short half-lives - after decay counting. 

If possible, continuous monitoring of the target stability during the activation is therefore preferable. This is often done by detecting the scattered beam particles from the target. For this purpose, a particle detector will be placed inside the activation chamber. The target degradation can then be derived from the yield of the scattered particles.

Problems with target thickness determination and target degradation can be avoided by using sufficiently thick targets in which the beam is completely stopped. In such a case, instead of the cross section the so-called thick target yield is measured directly. Applying fine energy steps, the cross sections can be deduced by differentiating the thick target yield function. If the thick target yield is measured in the Gamow-window for a given astrophysical process, then the astrophysical reaction rate can be derived directly from the yield values. Details of a thick target yield measurement and the related formulae can be found in ref.\,\cite{Gyurky2014112}.

\subsubsection{Gas targets}
\label{sec:gastarget}

Almost all considerations above are related to solid targets. In some cases, however, the application of a gas target for an activation experiment may be necessary or advantageous. In the case of noble gases, for instance, besides the often difficult-to-characterize implanted targets, a gas target is the only option. An example is the $^3$He($\alpha,\gamma$)$^7$Be reaction \cite{Bordeanu20131,PhysRevLett.48.1664,PhysRevC.27.11,1983ZPhyA.310...91V,PhysRevLett.93.262503,PhysRevLett.97.122502,PhysRevC.75.035805,PhysRevC.76.055801,PhysRevC.86.032801}, which involves two noble gas isotopes and necessitates the application of a gas target. A gas target was also needed for the study of the $\gamma$-process reaction $^{124}$Xe($\alpha,\gamma$)$^{128}$Ba \cite{PhysRevC.94.045801} as well as for the (n,$\gamma$) reactions on the Ne \cite{1991ApJ...379..420B,BEER2002239} 
and Xe isotopes \cite{1991ApJ...375..823B}. Other examples are reactions in inverse kinematics that involve proton or alpha-induced reactions using Hydrogen or Helium targets. For example, the $^{40}$Ca($\alpha,\gamma$)$^{44}$Ti reaction was studied in inverse kinematics with $^{44}$Ti counting using AMS \cite{PFA03,NPA06}.

Gas targets can be windowless (extended \cite{KOLLE1999160,COSTANTINI2008144} or gas jet \cite{Rapagnani2017217}) or gas cell type \cite{BORDEANU2012220}. In a cell the gas is confined between either two thin foils (where the beam passes through both foils) or one thin foil and the beam stop. For neutron activation high-pressure cells of aluminum, stainless steel or titanium have been used. Windowless gas targets are necessary at low bombarding energies when the energy loss and straggling would be too much even in the thinnest possible entrance window (the window would completely stop the beam or the energy of the beam after passing through the foil would be highly uncertain). Windowless configurations are also preferred for in-beam experiments where the prompt radiation emitted from the reactions taking place in the window could cause disturbing background. This latter issue is not of concern in activation experiments as disturbing activity originating from the window can easily be avoided (see sect.\,\ref{sec:background}).

The number of target atoms can be determined in a gas target experiment typically more precisely than for solid targets by measuring the gas pressure and temperature and knowing the physical length of the gas cell or chamber. For an extended gas target, the pressure profile along the length of the chamber must be investigated while the thickness of a jet target is typically measured by elastic scattering or nuclear reactions. Possible impurities present in the target gas must also be identified. If a high intensity beam passes through the gas, local heating may result in the reduction of the density and therefore a thinning of the target. A detailed study of the latter two effects can be found in ref.\,\cite{MARTA2006727}.

The radioactive reaction products created in the gas must be collected in a suitable catcher. The catcher can be the beam stop closing the gas volume on the downstream side, the foil closing a gas cell, or a separate foil placed inside the gas at a suitable place. Taking into account the reaction kinematics and the energy loss and straggling, it is important to guarantee that the reaction products can reach the catcher with high enough energy in order to be implanted deep enough into the catcher. Simulations using the actual geometry of the setup may be necessary \cite{PhysRevC.94.045801}.

\subsection{Neutron-induced reactions}
\label{sec:neutroninduced}

In contrast to activations with charged particles, measurements of neutron-induced reactions are limited by the neutron beam intensity. Neutron fluxes are
typically several orders of magnitude smaller than the intensities of proton or 
$\alpha$ beams. This difficulty is partly compensated by the longer 
range of neutrons in matter so that much thicker samples can be used
in neutron activations. This section deals with the role of neutron-induced reactions in nuclear astrophysics
and the possibility to mimic stellar neutron spectra in the laboratory for the 
corresponding cross section measurements. In particular, the concept of using 
quasi-stellar neutron spectra for activation measurements turned out to be a 
very efficient and comparably simple way of obtaining a wealth of ($n,\gamma$)
cross section data for nucleosynthesis studies in Red Giant stars.

 \subsubsection{Astrophysical scenarios and laboratory approaches \label{sec:neutroninduced.1}}

More than 95\% of the abundances of elements above Fe are the result of neutron-capture nucleosynthesis during stellar evolution (s process) and during 
some kind of explosive event, e.g. a final supernova or the merger of two
neutron stars (r-process). The s-process scenarios are related to the 
advanced evolutionary stages of shell-He and shell-C burning and are
characterized by temperature and neutron density regimes ranging from
0.1 to 1 GK and 10$^7$ to 10$^{10}$ neutrons/cm$^3$, respectively \cite{RevModPhys.83.157}.
In the explosive r-process environments temperatures and neutron 
densities are much higher, reaching 2-3 GK and more than 10$^{20}$
neutrons/cm$^3$. These parameters imply typical neutron capture
times of weeks to years inside the stars and of milliseconds in explosive
events, much longer or much shorter than average beta decay times, which are typically ranging between minutes and hours.

Accordingly, the s-process reaction path follows the valley of beta stability
by a sequence of ($n, \gamma$) reactions on stable or long-lived isotopes,
whereas the r process exhibits a complex reaction network of very short-lived
nuclei far from the line of stability. Experimental efforts are, therefore, concentrated 
on cross section measurements for the s-process, where data are needed at
keV neutron energies according to the temperatures mentioned above.
Free neutrons in stars are essentially provided by ($\alpha, n$) reactions 
on $^{13}$C and $^{22}$Ne during the helium burning phases of stellar evolution. 

In the dense stellar plasma neutrons are quickly thermalized and follow a 
Maxwell-Boltzmann energy distribution. The effective stellar ($n, \gamma$) 
cross sections are defined as Maxwellian averaged cross sections (MACS) \cite{RevModPhys.83.157}
by averaging the energy-dependent cross section over that spectrum,
\begin{equation}\label{eq:macs}
\langle \sigma \rangle_{kT}=\frac{2}{\sqrt{\pi}}\frac{\int_0^\infty 
  \sigma(E_n)~E_n~{\rm e}^{-E_n/kT}~dE_n}{\int_0^\infty 
  E_n~{\rm e}^{-E_n/kT}~dE_n}.
\end{equation}
To cover the full range of s-process temperatures, the cross sections 
$\sigma(E_n)$ are needed as a function of neutron energy from about
 $0.1 \leq E_n \leq 500$ keV. Such data are usually obtained in 
time-of-flight (TOF) measurements at pulsed neutron sources. 

Instead of evaluating the MACS via Eq. \ref{eq:macs}, activation in 
quasi-stellar neutron spectra offers an important alternative that allows 
one to determine the MACS values directly from the induced activity \cite{RevModPhys.83.157}. 

\subsubsection{Activation in quasi-stellar neutron spectra  \label{sec:neutroninduced.2}}

Apart from the fact that the method is restricted to cases, where neutron 
capture produces an unstable nucleus, activation in a quasi-stellar neutron 
spectrum has a number of appealing features.
\begin{itemize}
\item Stellar neutron spectra can be very well approximated under 
laboratory conditions so that MACS measurements can be immediately
obtained by irradiation and subsequent determination of the induced 
activity.
\item Technically, the method is comparably simple and can be performed 
at small electrostatic accelerators with standard equipment for $\gamma$ 
spectroscopy.
\item The sensitivity is orders of magnitude better than for TOF experiments,
 because the accelerator can be operated in DC mode 
and because the sample can be placed directly at the neutron production 
target in the highest possible neutron flux. This feature opens opportunities 
for measurements on sub-$\mu$g samples and on rare, even unstable isotopes, an 
important advantage if one deals with radioactive materials. 
\item In most cases the induced activity can be measured via the $\gamma$
decay of the product nucleus. This implies favorable signal/background
ratios and unambiguous identification of the reaction products. The
excellent selectivity achieved in this way can often be used to study 
more than one reaction in a single irradiation, either by using 
elemental samples of natural composition or suited chemical 
compounds.  
\item In the case of long-lived reaction products, direct atom counting through accelerator mass spectrometry can be applied (see sec. \ref{sec:AMS}). This method is complementary to decay counting.
\end{itemize}

So far, experimental neutron spectra, which simulate the energy dependence of the denominator of Eq.\,\ref{eq:macs} have been produced  by three reactions. The $^7$Li($p, n$)$^7$Be reaction provides a
spectrum similar to a distribution for a thermal energy of $kT$ = 25 keV 
\cite{BeK80,RaK88} very close to the 23 keV effective thermal energy in He 
shell flashes of low mass AGB stars, where neutrons are produced via the 
$^{22}$Ne($\alpha, n$)$^{25}$Mg reaction. Alternative possibilities are
quasi-stellar spectra for $kT=5$\,keV  \cite{HDJ05} and 52\,keV \cite{KNA87}
that can be obtained with ($p, n$) reactions on $^{18}$O and $^3$H,
respectively. The spectrum of 5\,keV is well suited to mimic the main 
neutron source in AGB stars, because the $^{13}$C($\alpha, n$) source 
operates at 8 keV thermal energy, whereas the spectrum of 52\,keV is
similar to the higher temperatures during shell-C burning in massive stars
($kT=90$ keV). More specific spectra can be obtained by the superposition of irradiations at different energies and sample positions as demonstrated in Ref. \cite{Reifarth2018}.

\begin{figure}
\includegraphics[width=8.5cm]{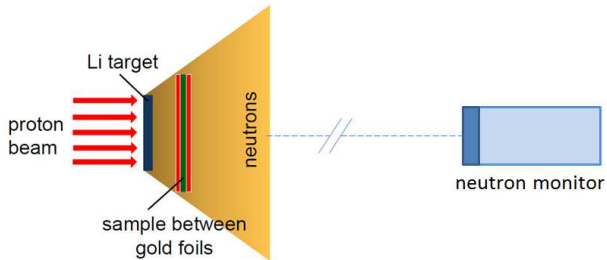}
\caption{\label{fig:actmacs} Schematic setup for activations in a quasi-stellar 
neutron spectrum. The energy of the primary proton beam is chosen 
such that the neutrons from $^7$Li($p, n$) reactions are kinematically 
collimated. The sample is sandwiched between gold foils for flux 
normalization, and a neutron monitor is used for recording the 
irradiation history.}
\end{figure}

Because the proton energies for producing these quasi-stellar spectra
are only slightly higher than the reaction thresholds, all neutrons 
are emitted in forward direction as illustrated schematically in fig. 
\ref{fig:actmacs}. The samples are placed such that they are exposed to the full spectrum, but very close to the target at distances of typically 1\,mm. The simultaneous activation of gold foils in front and back of the samples are used to determine the neutron flux via the well-known ($n,\gamma$) cross section of ${197}$Au. 
The setup includes a neutron monitor at some distance from the 
source for recording the neutron intensity during the irradiation. 
This information serves for off-line corrections of intensity 
variations due to fluctuations of the proton beam or to a degradation 
of the target \cite{BeK80}, an aspect that is important if the half-life 
of the induced activity is comparable to the irradiation time. 

With the available proton beam currents of electrostatic accelerators of up to 100\,$\mu$A \cite{URS09} it is possible to produce maximum yields of 10$^9$, 10$^8$, and 10$^5$ neutrons per second via ($p,n$) reactions on $^7$Li, $^3$H, and ${18}$O, respectively. These values are orders of magnitude higher than obtainable in TOF experiments. For comparison, the highest fluxes reached at the 
most intense TOF facilities LANSCE \cite{LBR90} and 
n\_TOF-EAR2 at CERN \cite{SBC17} are about 5$\times$10$^5$ 
s$^{-1}$. A further increase by more than an order of magnitude 
in beam power and, correspondingly in neutron flux, has been 
gained in activation measurements at the SARAF facility \cite{PTF14}.

Thanks to the high neutron flux, activation represents the most 
sensitive method for ($n, \gamma$) measurements in the astrophysically 
relevant energy range. This feature provides unique possibilities to 
determine the very small MACSs of neutron poisons, of abundant 
light isotopes, and of neutron magic nuclei. Moreover, the excellent
sensitivity is of fundamental importance in measurements on very 
small samples, be it because the sample material is extremely rare
as in the case of $^{60}$Fe or comparably short-lived. The latter aspect 
is crucial for the determination of the MACSs of unstable isotopes, 
which give rise to local branchings in the $s$-process path by the 
competition between neutron capture and $\beta^-$ decay as in case of ${147}$Pm discussed below. The
branchings are most interesting because the evolving abundance 
pattern carries information on neutron flux, temperature and pressure 
in the stellar plasma (see Ref. \cite{KGB11} for details). In most cases,
TOF measurements on unstable branch point isotopes are challenged 
by the background due to the sample activity or because sufficient 
amounts of isotopically pure samples are unavailable. 

Another advantage of the activation method is that it is insensitive 
to the reaction mechanism. In particular, it includes the contributions 
from Direct Radiative Capture (DRC), where the neutron is captured 
directly into a bound state. This component contributes substantially 
to the ($n, \gamma$) cross sections of light nuclei, but could not be 
determined in TOF measurements so far. 
 
Likewise, the determination of partial cross sections leading to the 
population of isomeric states, which is very difficult in TOF experiments, 
can easily be performed by activation \cite{HWD08}. 

Certain limits to the activation method are set by the half-life of the 
product nuclei. Long half-lives imply low induced activities, which are
then very difficult to quantify accurately. In favorable cases, this problem 
can be circumvented by means of the AMS technique discussed in Sec. 
\ref{sec:AMS}. In case of short half-lives, saturation effects are restricting the 
induced activity at a low level, which is then further reduced by 
substantial decay between irradiation and activity counting. By repeated 
cyclic activation, this limit can be pushed to a few seconds \cite{BRW94}.

\subsubsection{Selected examples \label{sec:neutroninduced_examples}}

The examples of the MACS measurements on $^{19}$F \cite{UHK07}, 
$^{60}$Fe \cite{URS09}, and $^{147}$Pm \cite{RAH03} are chosen 
because they illustrate how even situations near the technical limits can 
be handled thanks to the excellent sensitivity of the activation method. 

\begin{figure}
\includegraphics[width=8.5cm]{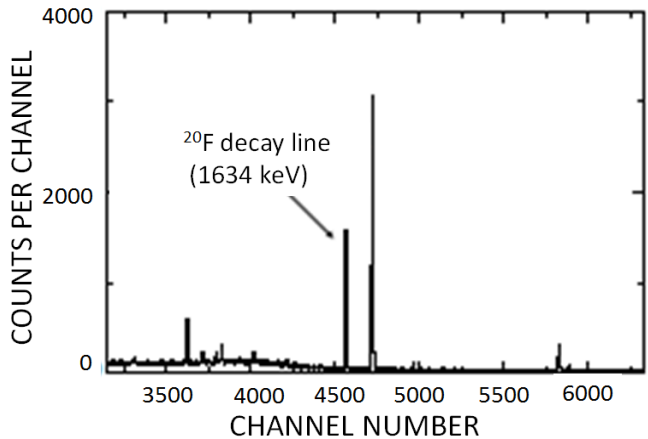}
\caption{\label{fig:19F} Cumulated $\gamma$ spectrum from the
cyclic activation of $^{19}$F. The $^{20}$F decay line stands clearly out 
of the background. The other lines are from activated materials surrounding 
the detector. (fig. from Ref. \cite{UHK07}).}
\end{figure}

The $^{19}$F MACS measurement \cite{UHK07} is challenging because of the relatively short half-life of 11\,s of the radioactive $^{20}$F isotope. Correspondingly, the irradiation time was limited to about 30 s 
to avoid critical saturation effects. In turn, the small MACS of $^{19}$F 
implied that not enough activity could be produced in this short period. 
In this case cyclic activations were performed using a pneumatic slide to 
transport the sample within 0.8 s from the irradiation position of the $^7$Li target to a heavily shielded HPGe detector at a distance of 50 cm, each cycle 
lasting for 60 s. During the counting interval, the proton beam was blocked 
to keep the background at a manageable level. The cumulated $\gamma$
spectrum in fig. \ref{fig:19F} illustrates that very clean conditions could be
obtained in spite of the experimental difficulties.

In the second case the  6 min half-life of $^{61}$Fe was sufficient for 
transporting the sample to a low-background laboratory for
the activity measurement. This activation was complicated by the small 
MACS of 5.7 mb and by the minute sample \cite{SND10} of only 
1.4 $\mu$g, which resulted in an extremely low activity per cycle and
required 47 repeated irradiations. The third experiment was performed 
with an even smaller sample of only 28 ng or $1.1\times10^{14}$ 
atoms in order to keep the $^{147}$Pm activity ($t_{1/2}$=2.6234\,$\pm$\,0.0002\,y) 
at a reasonable value of 1\,MBq. In both measurements a compact arrangement of 
two high-efficiency HPGe Clover detectors was required to identify the
weak $\gamma$ signals from the activation. This setup is described 
in the following Sec. \ref{sec:secondphase}. 

In connection with recent observations of terrestrial ${60}$Fe the yet unmeasured MACS of ${59}$Fe became an important issue. $^{60}$Fe is mostly produced in the late evolutionary stages of massive stars
and is distributed in the interstellar medium by subsequent supernova
explosions \cite{LiC06}. Minute traces of $^{60}$Fe have also been discovered in 
deep sea sediments pointing to nearby supernovae within the past few Myr 
\cite{WFK16}. To provide the complete link between the amount of 
$^{60}$Fe produced and the traces found on earth one has to know the MACS
of the short-lived $^{59}$Fe ($t_{1/2}$ = 44.503\,$\pm$\,0.006 d). In view of the inconveniently 
short half-life this measurement appears only feasible if the MACS can be inferred
indirectly. This could be obtained using double neutron capture sequence 
$^{58}$Fe($n, \gamma$)($n, \gamma$)$^{60}$Fe, irradiating stable 
$^{58}$Fe in an intense quasi-stellar spectrum for $kT=25$ keV and detecting 
the final product $^{60}$Fe via AMS. However, this venture represents a really 
big challenge and requires extremely high neutron densities of the order of 10$^{12}$ s$^{-1}$ that may,
hopefully, be reached once the future FRANZ facility \cite{ABB16} will be fully 
operational.

\begin{figure}[ht]
 \includegraphics[width=8.5cm]{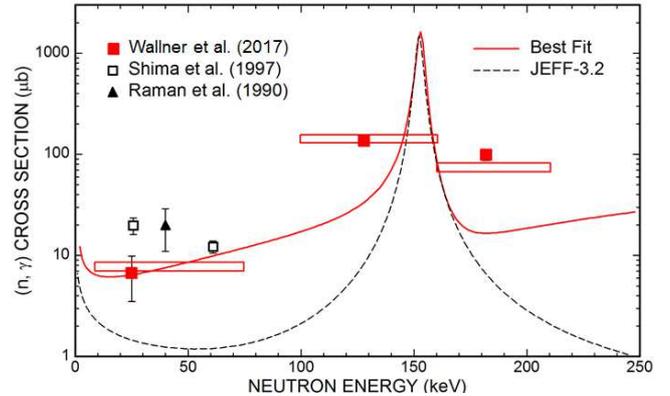}
 \caption{The $^{13}$C($n, \gamma$)$^{14}$C cross section 
between 1\,keV and 300\,keV \cite{WBB16}. The red solid line represents a best-fit cross section that describes the experimental results (black squares; open red boxes indicate the FWHM in neutron energy) and the resonance at 154 keV consistently. The p- and d-wave 
components of the DRC contribution were neglected in the JEFF 3.2 
evaluation \cite{JEF14}, but contribute substantially in the astrophysically 
relevant region below the resonance at 152.4 keV. (Figure from Ref. \cite{WBB16}). 
\label{fig:sigma13C}}
\end{figure}

Sometimes MACS measurements in a quasi-stellar spectrum need to 
be complemented by additional activations at higher energies. This
is illustrated at the example of the $^{13}$C($n, \gamma$) reaction \cite{WBB16}. Fig. \ref{fig:sigma13C} shows how the p- and d-wave components of the DRC contribution could be quantified at the relevant stellar energies around 25\,keV by means of two additional activations between 100\,keV and 200\,keV, just below and above the 154-keV resonance.

A full collection of the many activation measurements in quasi-stellar neutron 
fields can be found in the KADoNiS compilation \cite{DPK10}.

\subsection{Gamma induced reactions}
\label{sec:gammainduced}

From the point of view of nuclear astrophysics, photon-induced experiments together with a general description of the experimental approaches have been summarized in the review paper of Mohr et al. \cite{Mohr2007}. In the present work a brief summary is provided on the astrophysical motivation of $\gamma$-induced reaction studies with the activation method together with a brief account of state-of-the art $\gamma$-sources and the experimental setups relevant to    
activation. A special emphasis is given to the upcoming ELI-NP facility \cite{0295-5075-117-2-28001} opening new possibilities for the $\gamma$-induced reaction studies.
Some examples will be provided as well.
 
 \subsubsection{Astrophysical motivation \label{secgammainduced.1}}

Laboratory studies of $\gamma$-induced reactions can be important either for astrophysical scenarios where $\gamma$-induced reactions are dominant, or to study radiative capture reactions where the direct study is difficult from the technical point of view. The astrophysical $\gamma$-process responsible mainly for the production of p-nuclei is clearly connected to a sequence of $\gamma$-induced reactions, therefore many experiments have been performed to study the nuclear physics background of p-process nucleosynthesis. A review of Rauscher et al \cite{0034-4885-76-6-066201}  summarized the astrophysical origin of the p-nuclei, the relevant reaction rates and reaction mechanisms, and in general the nuclear physics aspects of the $\gamma$-process. An indirect study of $(n,\gamma)$ reactions for the  s-process through the inverse $(\gamma, n)$ reaction is another example for using $\gamma$-induced reactions \cite{PhysRevC.73.025804}.

In an astrophysical scenario, i.e. a given layer of a supernova explosion, the photon density at temperature $T$ is \cite{Mohr2007}:
\begin{equation}
n_\gamma(E,T) = 
  \left( \frac{1}{\pi} \right)^2 \,
  \left( \frac{1}{\hbar c} \right)^3 \,
  \frac{E^2}{\exp{(E/kT)} - 1}
\label{eq:planck}
\end{equation}
and the stellar reaction rate of a $\gamma$-induced reaction ($\gamma$,x) is:
\begin{equation}
\lambda^\ast(T) = 
 \int_0^\infty 
  c \,\, n_\gamma(E,T) \,\, \sigma^\ast_{(\gamma,x)}(E) \,\, dE
\label{eq:rate}
\end{equation}
It is important to note that $\sigma^\ast_{(\gamma,x)}(E)$ is the cross section under stellar conditions, that can differ in some cases drastically from the laboratory value where the target is always in the ground state. This is why in many cases the reverse charged particle induced reactions are studied instead of the $\gamma$-induced ones. A wide range of activation experiments have been performed in that way.

\begin{figure}[ht]
 \includegraphics[width=8.5cm]{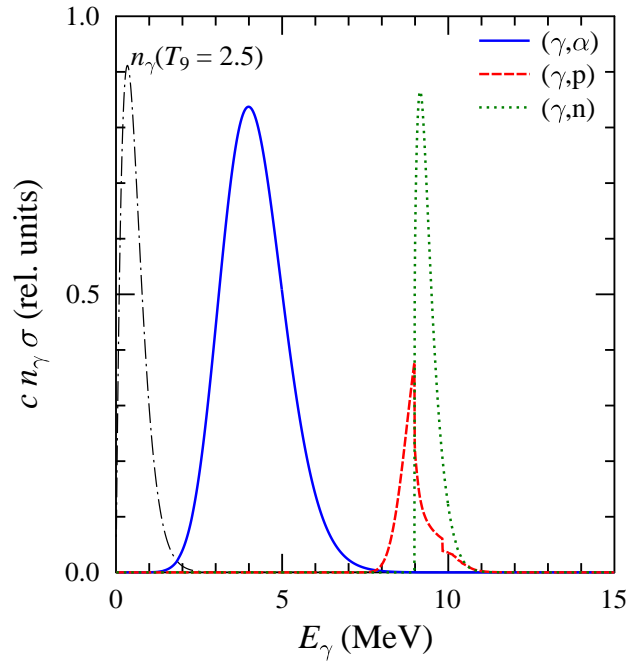}
 \caption{Relative yields for the $^{148}$Gd($\gamma$,x) reactions to demonstrate the position of the astrophysically relevant energy region for the reactions. Note that the yields are scaled individually for better visibility. The dash-dotted line represents the Planck distribution of the photon energy. See text for details.
\label{fig:integ_g-x}}
\end{figure}

As an example of the astrophysically important energy region for the $\gamma$-process, we show in fig.\,\ref{fig:integ_g-x} the above
integrand for $^{148}$Gd$(\gamma,\alpha)$, $^{148}$Gd$(\gamma$,p), $^{148}$Gd$(\gamma$,n) at $T_9$=2.5 in relative
units. The cross sections have been taken from the TALYS code in the capture channel and have then been converted to the  $^{148}$Gd($\gamma$,x) reactions, so it
is relevant only for the laboratory yields. In addition, the energy region (Gamow window) depends also on the mass of the nucleus and the site temperature. Consequently, $(\gamma$,n) studies need $\gamma$-beams ranging in energy from 1\,MeV up to 10\,MeV.

 \subsubsection{Relevant $\gamma$-sources \label{secgammainduced.2}}

There are numerous ways to produce high energy photons with high intensities, and worldwide there are dedicated facilities providing $\gamma$-beams. While a wide range of sources includes $\gamma$-ray production using thermal neutron capture and positron annihilation in flight, in the present paper we discuss those facilities where the activation method has been used recently or is planned to be used for astrophysical purposes. Those facilities use either bremsstrahlung radiation or laser Compton scattering to produce high energy $\gamma$-rays.  The basic performance parameters of the systems are the beam intensity and the energy resolution.

It has to be noted that while tagging is used in many setups to improve the energy resolution of the system (see e.g. ref.\,\cite{SAVRAN2010232}), this cannot be used for the activation experiments, so we omit the discussion about the tagging procedure here.

The bremsstrahlung facilities consist typically of a high energy, high intensity electron accelerator and a radiator target, where the electron beam slows down and a continuous energy $\gamma$-spectrum is released. S-DALINAC, Darmstadt \cite{SONNABEND20116} and ELBE, Dresden \cite{TEICHERT2003354} are facilities where such astrophysics-related activation experiments have been performed. Cross sections of $\gamma$-induced reactions can be determined at bremsstrahlung facilities basically with two methods. In the first one, yield differences are measured and unfolded at different electron energies with the corresponding continuous $\gamma$-spectra \cite{RevModPhys.47.713}. In the second method, instead of individual activations, a superposition of bremsstrahlung spectra is designed in a way that a $\gamma$-field of the astrophysical scenario (or at least its high energy domain) is approximated \cite{2000PhLB..488..127M}.

In both methods, the crucial part is the determination of the absolute $\gamma$-yield, and the electron energy.

A further methodology for $\gamma$-beam production is the use of laser Compton scattering, i.e. Compton scattering of a laser photon with a relativistic electron. In contrast to the bremsstrahlung sources this kind of facilities provide quasi-monoenergetic photon beams of variable energies. A summary of the technology and recent developments is given in \cite{HAJIMA201635} and the HIGS (High Intensity Gamma-Ray Source) facility is described in details in  \cite{WELLER2009257}.

Since the Nuclear Physics pillar of the Extreme Light Infrastructure (ELI-NP) in Romania is being commissioned, we will describe this facility as an example of a $\gamma$-beam facility based on laser Compton scattering.

The ELI-NP $\gamma$-beam system (GBS) \cite{2014arXiv1407.3669A} will be superior to the laboratories which are operational at present in terms of beam intensity and bandwidth (see table 1 in ref~\cite{Filipescu2015} and references therein). The facility will deliver almost fully polarized, narrow-bandwidth, high-brilliance $\gamma$-beams in the energy range between 200 keV and 19.5 MeV, which will be produced via inverse Compton backscattering of laser photons off relativistic electrons.  The time structure of the $\gamma$-beams will reflect the radiofrequency (RF) pulsing of the electron accelerator working at a repetition rate of 100 Hz. Each RF pulse will contain 32 electron bunches with an electric charge of 250 pC, separated by 16 ns. The J-class Yb:YAG lasers will deliver 515 nm green light and will operate at 100 Hz. A laser re-circulator will ensure the interaction with the train of 32 microbunches \cite{PhysRevSTAB.17.033501}.  The parameters of the ELI-NP $\gamma$-beams are summarized in Table\,\ref{tab:elibeams}. 

\begin{table*}
\center
\caption{\label{tab:elibeams} Parameters of the ELI-NP $\gamma$-beams \cite{2014arXiv1407.3669A}}
\begin{tabular}{ll}
GBS parameter	&	Value	\\
\hline
Energy (MeV)	&	0.2 –- 19.5	\\
Spectral density (10$^4$ photons/s/eV)	&	0.8 –- 4	\\
Bandwidth (rms)	&	$\leq$ 0.5	\\
Photons/shot within FWHM	&	$\leq$ 2.6$\cdot$10$^5$	\\
Photons/s within FWHM	&	$\leq$ 8.3$\cdot$10$^8$	\\
$\gamma$ beam rms size at interaction [$\mu$m]	&	10 -– 30	\\
$\Gamma$ beam divergence [$\mu$rad]	&	25 -– 200	\\
Linear polarization [\%]	&	$>$ 95	\\
Repetition rate [Hz]	&	100	\\
Pulses per macropulse	&	32	\\
Separation of microbunches [ns]	&	16	\\
Length of micropulse [ps]	&	0.7 -– 1.5	\\
\hline
\end{tabular}
\end{table*}

\begin{figure}[ht]
 \includegraphics[width=8.5cm]{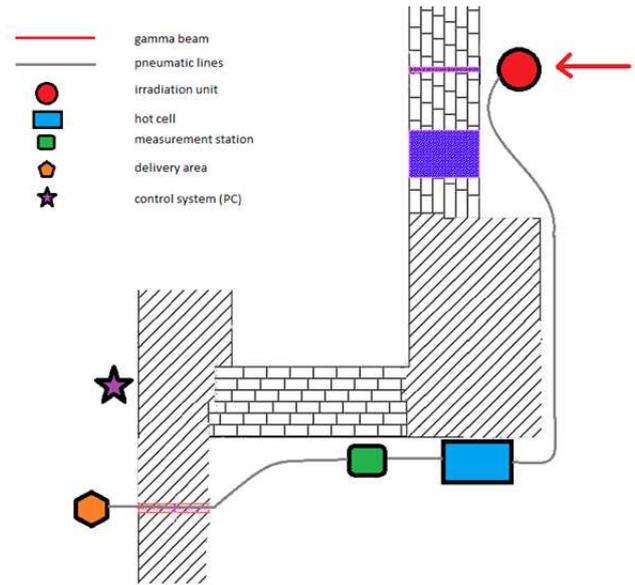}
 \caption{The irradiation facility for the ELI-NP $\gamma$-source. See text for details.
\label{fig:ELI}}
\end{figure}

At the ELI-NP GBS it will be possible to perform activation experiments. For this purpose a dedicated irradiation station \cite{bobeica2016} is under construction. It is designed for irradiation of various solid targets with an intense $\gamma$-beam. The system has to be able to host several solid targets, automatically load a target and position it for irradiation. After the irradiation, the target need to be moved from the target position and transferred to the target measurement station by e.g. a pneumatic transport system. All these operations are to be done remotely via a computer control system. For achieving an optimal irradiation of the target, the alignment of the irradiation unit will be done remotely via stepper motors with an accuracy of $\pm$0.1mm. To control the alignment of the target as well as the beam hitting point on the target itself, a CCD camera will be part of the alignment system. During the irradiation, the target has to be also aligned in the horizontal plane with very high accuracy. The alignment system will keep a correct angle alignment between the symmetry axis of the target cylinder and the beam axis within $\pm 0.5 ^\circ$. After the irradiation process, targets are transported via a devoted mechanical system to the measurement station equipped with Pb shielded and efficiency calibrated HPGe detectors. The setup for activation experiments at ELI-NP is shown schematically in fig. \ref{fig:ELI}.

 \subsubsection{Activation experiments \label{secgammainduced.3}}

In this section, further information about some $\gamma$-beam facilities is given where the activation method has been used to measure $\gamma$-induced cross sections.
It is not possible to cover all the experiments in the field, instead, 
selected instruments with selected experimental approaches are
reported with limited details. The reader can find the full
experimental descriptions in the relevant references.

Photoactivation technology has been used to measure partial photoneutron 
cross sections
on $^{181}$Ta($\gamma$,n)$^{180}$Ta, since partial cross sections for the isomeric state  can probe
the nuclear level density of $^{180}$Ta. 
In this experiment, the total cross section was determined by
direct neutron counting, while the ground state cross section 
by photoactivation \cite{PhysRevLett.96.192501}.
This experiment was carried out in Japan, 
at the LCS (Laser Compton Scattering) 
beamline of the National Institute of Advanced 
Industrial Science and Technology (AIST).

A wide range of experiments on the direct determination of $(\gamma,n)$
cross sections with activation has been carried out at the 
S-DALINAC \cite{SONNABEND20116} and ELBE \cite{TEICHERT2003354} facilities.
Those experiments can reveal the importance of nuclear data for  heavy element nucleosynthesis. Results on systematic 
investigations using
 S-DALINAC on various $(\gamma,n)$ reactions are now available \cite{PhysRevC.73.025804,PhysRevC.63.055802,PhysRevC.70.035802,PhysRevC.77.015803,PhysRevC.79.055807}.

At ELBE, a pneumatic delivery system  (RABBIT) has been designed 
to determine the activity of
short lived residual isotopes. The studies at ELBE helped to understand 
the dipole
strength and modified photon strength parameters could be suggested and
compared to experimental data \cite{PhysRevC.78.055802,PhysRevC.81.055806,Erhard2006,PhysRevC.81.034319}.

The first photodisintegration cross sections determined at a commercial medical linear accelerator were reported 
recently \cite{vagena2017} aiming at ($\gamma$,n) reactions on various Dy isotopes.
Since those accelerators are widely 
spread, this could be a very useful tool to carry out similar 
nuclear astrophysics studies at medical centers in the future.



\section{Second phase of an activation experiment: determination of the number of produced nuclei}
\label{sec:secondphase}

After the irradiations, the number of radioactive nuclei produced must be determined by the measurement of some decay radiation. If the half-life of the reaction product is short (typically less than a few minutes), then either a fast delivery system is needed which transports the target from the activation chamber to the counting facility (see e.g. ref.\,\cite{0954-3899-35-1-014036}) or a detector must be placed next to the target chamber which allows to measure the activity before radioactive nuclei have decayed. In the case of short half-lives cyclic activation is often needed in order to collect enough statistics.

In most cases, however, the half-life of the reaction product is long enough so that the target can be removed from the activation chamber and transported to a detector where the decay can be measured. With only very few exceptions the radioactive reaction products undergo $\beta$-decay. All three types of $\beta$-decay ($\beta^-$, $\beta^+$ and electron capture) are encountered. The $\beta$-decay very often leaves the daughter nucleus in an excited state and therefore the decay can be followed by $\gamma$-emission. 

Since $\gamma$-detection has some clear advantages compared to $\beta$-detection (lower self-absorption of $\gamma$-rays in the target compared to $\beta$-particles, well defined $\gamma$ transition energies as opposed to continuous $\beta$-spectra), in the overwhelming majority of cases, $\gamma$-detection is used in nuclear astrophysics activation experiments. In the next subsection the experimental aspects of $\gamma$-detection are detailed. Other cases will be discussed in sect.\,\ref{sec:rarecases}.

\subsection{Gamma-detection}
\label{sec:gammadetection}

The de-excitation of the daughter nucleus populated in a $\beta$-decay involves the emission of characteristic $\gamma$-rays of well-defined energy. Not too far from the stable isotopes the energies and relative intensities of these characteristic $\gamma$-rays are typically well known (see, however, sect.\,\ref{sec:decaypar}), hence the number of produced isotopes can be determined from the detection of $\gamma$-rays. HPGe (High Purity Germanium) detectors \cite{2008pgrs.bookG} have excellent energy resolution allowing to discriminate different isotopes or elements present in the target.

\begin{figure}
	\includegraphics[angle=270,width=\columnwidth]{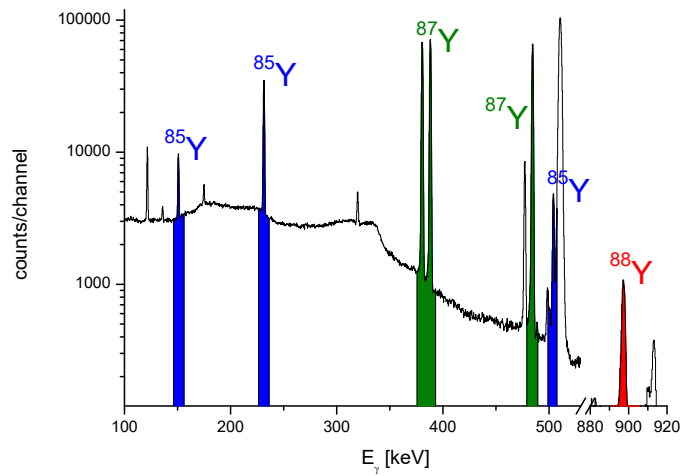}%
	\caption{Activation $\gamma$-spectrum taken on a natural Sr target irradiated with a 3\,MeV proton beam \cite{PhysRevC.64.065803}. The peaks belonging to the different reaction products are colour-coded.}
	\label{fig:Sr_spectrum}
\end{figure}

As an example, fig.\,\ref{fig:Sr_spectrum} shows an activation $\gamma$-spectrum taken on a natural Sr target after irradiation with a 3\,MeV proton beam. Proton capture on three stable Sr isotopes ($^{84,86,87}$Sr) leads to radioactive Y isotopes ($^{85,87,88}$Y). The decay of the three isotopes can easily be identified in the $\gamma$-spectrum shown in the figure. $^{85}$Y and $^{87}$Y have long-lived isomeric states. Owing to the different $\gamma$-radiations, the decay of the ground and isomeric states could be measured separately and partial cross sections leading to these states could therefore be derived \cite{PhysRevC.64.065803}. 

If the half-lives of the studied reaction products differ significantly, the timing of the decay counting may also help to identify different produced isotopes. An example can be found in ref.\,\cite{PhysRevC.90.052801}.

For an absolute determination of the cross section the most important quantity of a $\gamma$-detector is the absolute $\gamma$-ray detection efficiency. Indeed, the peak area $A$ in the $\gamma$-spectrum from a given transition is related to the total number of decays $N_{decay}$ in the counting interval by the following simple relation \cite{Kugler}:

\begin{equation}
	A=N_{decay} \cdot \epsilon \cdot \eta
\end{equation}
where $\eta$ is the relative intensity of the transition in question (i.e. the ratio of the emitted $\gamma$-rays to the number of decays) and $\epsilon$ is the absolute efficiency of the detector at the given $\gamma$-energy (i.e. the ratio of the number of events in the full energy peak to the number of $\gamma$-rays emitted). Since radioactive sources emit $\gamma$-rays with isotropic angular distribution, no term is needed in the above formula for the angular distribution (see, however, the case of coincidence technique below). Based on the methods discussed in the next paragraphs, the $\gamma$-detection efficiency can typically be determined to a precision of a few percent.

As opposed to in-beam $\gamma$-spectroscopy, in activation usually only low $\gamma$-energies (typically below 2\,MeV) are encountered. In this energy range the detector efficiency can be measured by commercially available or custom-made radioactive calibration sources. 

Low cross sections typically lead to low activities in the targets. Therefore, in order to maximize the detection efficiency, close detection geometries and large volume detectors are required. In such a case, the so-called true coincidence summing effect may complicate the determination of the efficiency and the measurement of the target activity. True coincidence summing occurs if two or more $\gamma$-quanta from the decay of a single nucleus reach the detector \cite{debertin1988gamma}. The magnitude of the effect does not depend on the source activity, but strongly depends on the counting geometry and becomes significant at the close distances typically needed in nuclear astrophysics. The summing effect influences both the efficiency determination (if calibration sources emitting multiple $\gamma$-radiations are used) and the counting of the actually produced isotopes. 

\begin{figure}
	\includegraphics[angle=270,width=\columnwidth]{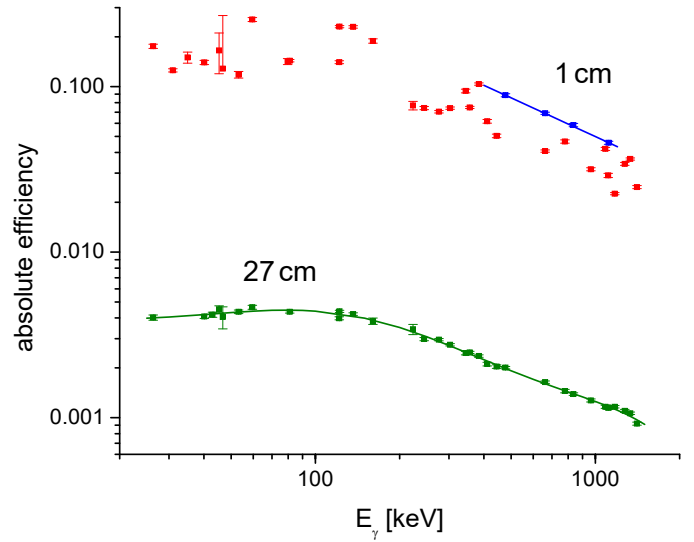}%
	\caption{Absolute efficiency of a HPGe detector measured in two geometries. See text for details.}
	\label{fig:deteff}
\end{figure}

To demonstrate the importance of the true coincidence summing effect, fig.\,\ref{fig:deteff} shows the absolute efficiency of a 100\,\% relative efficiency (relative to a 3'' $\times$ 3'' NaI scintillator detector) HPGe detector measured in two different geometries. Many calibrated radioactive sources were used for these measurements, some of them emitting only one single $\gamma$-ray ($^{7}$Be, $^{54}$Mn, $^{65}$Zn, $^{137}$Cs) and some of them emitting multiple $\gamma$-rays ($^{22}$Na, $^{57}$Co, $^{60}$Co, $^{133}$Ba, $^{152}$Eu, $^{241}$Am). 

One set of measurements was carried out mounting the target 27\,cm far from the detector while in the second measurement the distance was reduced to 1\,cm. As the diameter of the Ge crystal in the detector is 80\,mm, the first can be considered as ``far'' geometry where the summing effect is negligible and the second one as ``close'' geometry introducing a pronounced summing effect. Indeed, the measured efficiency as a function of energy in the far geometry can be fitted by a smooth curve indicating negligible summing. On the contrary, the points measured in close geometry show a large scatter caused by summing. Consequently, thees points cannot be used directly to obtain an efficiency function. The four points (shown in blue in fig.\,\ref{fig:deteff}) provided by single line sources, on the other hand, follow the usual power law energy dependence of the efficiency, as it is shown by the fit (blue line). This indicates that in a close geometry single line sources are strongly preferred for the efficiency determination. This, however, does not solve the problem related to the summing effect of the studied isotope itself.

Of course there are mathematical methods and computer codes which can be used to carry out coincidence summing corrections for obtaining reliable efficiencies and activities. Examples are GESPECOR \cite{GESPECOR} and TrueCoinc \cite{TrueCoinc}. For a precise summing correction, however, detailed information about the source and the detectors geometry is needed and typically the accuracy of the determined efficiency and activity remains inferior when compared with far geometry measurements.

A more precise method to avoid the coincidence summing problem can be the application of the two distance technique. In this method, a strong source containing the studied isotope is produced by an appropriate nuclear reaction. This can be a reaction leading to the same isotope but having a higher cross section \cite{0954-3899-37-11-115201}, or the studied reaction itself at higher energy where the cross section is larger \cite{PhysRevC.85.025804}. This source is measured in both far and close geometries. In far geometry the absolute efficiency is measured precisely with calibration sources. Then, taking into account the decay between the two measurements, an effective efficiency conversion factor is calculated for all studied $\gamma$-transitions from the comparison of the two measurements. This conversion factor contains the actual efficiency difference between the two geometries as well as the effect of the summing. The lower activity sources of the same isotope can then be measured in the close geometry and using the conversion factor, there is no need for the direct efficiency determination.

For a reliable cross section measurement the unambiguous identification of the reaction products and the reactions themselves is crucial. If the decay of the reaction product is followed by a single $\gamma$-ray emission, it is crucial to avoid any contamination in the studied peak. Such a contamination can be caused by another radionuclide produced on a target contaminant. If the half-life is short enough, the decrease of the activity can be followed. If the decay deviates from the decay law of the studied isotope, this may indicate the presence of a longer or shorter lived contaminant nuclei. If the counting statistics allows, shorter-lived contamination can be avoided by increasing the waiting time $t_w$, while the contribution of longer lived contaminants can be taken into account based on a later measurement when the studied isotope has already decayed. If the studied isotope emits multiple $\gamma$-rays, the comparison of activities determined from the different radiations can also reveal contamination.

It is also important to make sure that the final isotope of the studied reaction cannot be co-produced by any other reaction. Indeed, if this were the case, the activity measurement would never be able to distinguish the different reactions. Although rather rare, it can happen that the product of a charged particle capture or $\gamma$-induced reaction can also be produced by neutron capture of another stable isotope which may be present in the target \cite{TICKNER201099}. If this is possible, target purity is crucial and test measurements must be carried out. 

If the $\gamma$-decay scheme of the daughter isotope involves cascade transitions, coincidence measurements become possible \cite{2002NuPhA.710..469O,a4f5e5a33acd43bfa952482c6f955508,PhysRevC.84.045808,Netterdon2013149,PhysRevC.90.065807,SCHOLZ2016247}. In such a case the target is typically put between two high efficiency $\gamma$-detectors in close geometry for recording two members of a $\gamma$-cascade in coincidence. With such a technique the background can be substantially reduced but the precise determination of the coincidence efficiency may become more complicated. Additionally, angular correlations between the two studied $\gamma$-rays must also be taken into account. This is often neglected, although in some cases a negligibly small correlation effect is proven by calculations \cite{PhysRevC.84.045808}.

The limited efficiency of HPGe detectors has triggered the development of 
multi-detector arrays. In particular, Clover detector systems have been
used in activation measurements, which are combining four coaxial 
n-type HPGe diodes in a common cryostat. The diodes are cut in a way 
that enables close packing with a Ge-Ge distance of only 0.2 mm. A 
common solution consists of four detectors, each 50 mm in diameter, 
and 70 mm in length. With a total active volume of $\sim470$ cm$^3$
such a detector reaches an absolute efficiency of about 7\% at 1 MeV 
$\gamma$ energy. 

\begin{figure}[ht]
 \includegraphics[width=7cm]{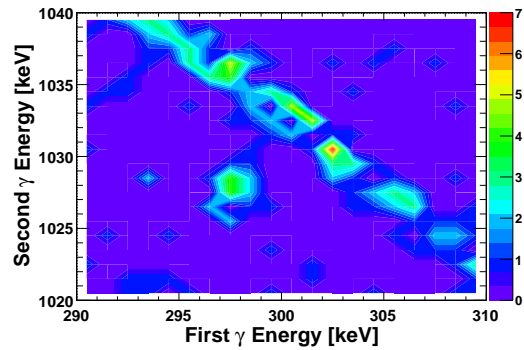}
 \caption{Coincidence spectra of the 298-1027 keV cascades in
the decay of $^{61}$Fe (green locus at the corresponding energies). The diagonal feature corresponds to events
with multiplicity two originating from scattering of the 1332
keV $^{60}$Co line. (fig. from Ref. \cite{URS09}). \label{fig:clovercascade}}
\end{figure}

Clover detectors have been used in the $^{60}$Fe activation experiment \cite{URS09} discussed in Sec.\,\ref{sec:neutroninduced_examples}. The use of such detectors has been needed because of the dominant background and the weak signal (stemming from the small amount of target material). The eight-fold segmentation of the adopted detectors allowed for the coincidence measurement of the $\gamma$-cascade.
By placing the sample between the Clovers in very 
close geometry, absolute peak efficiencies of 26\% and 10\% could be 
achieved for the cascade transitions at 298\,keV and 1027\,keV, respectively. 
Fig. \ref{fig:clovercascade} shows that a 
nearly background free spectrum could be obtained in this way, although 
with reduced counting statistics. 

\subsection{Other detection methods for activation measurements}
\label{sec:rarecases}

Gamma-ray detection is by far the most commonly applied detection method for activation cross section measurements in nuclear astrophysics. In some cases, however, this is not possible and other decay signatures must be detected. In this section four examples for such cases are shown.

\subsubsection{Alpha decay}
\label{sec:alphadecay}

In the overwhelming majority of cases in nuclear astrophysics, radioactive reaction products decay by $\beta$-decay. Only for the heaviest elements and in some special cases are $\alpha$-emitting isotopes produced. Since $\alpha$-decay is typically either not followed by any $\gamma$-emission, or produces only very low energy and low intensity $\gamma$-rays, the detection of the emitted $\alpha$ particles is preferable. 

Alpha particles can be detected with high (in ideal cases up to 100\,\%) efficiency, but since alphas have high energy loss in matter, the self attenuation of the target layer can be significant and limits the spectroscopic information that can be obtained. Reducing the background is important as well as avoiding contaminations. Two examples of nuclear astrophysics motivated $\alpha$-detection activation experiments are mentioned below.

One of the first reactions studied directly for the astrophysical $\gamma$-process was the $^{144}$Sm($\alpha,\gamma$)$^{148}$Gd reaction. As the heavy $^{148}$Gd lies two neutrons above the N\,=\,82 closed neutron shell, it decays back to the neutron magic $^{144}$Sm by $\alpha$-emission. $^{148}$Gd has a half-life of about 75 years and the decay is not followed by $\gamma$-emission. This reaction was studied by activation \cite{1998A&A.333.1112S}. The $\alpha$-particles emitted by $^{148}$Gd were detected either by a Si surface barrier detector, or in the case of lower activity samples by etched track detectors. The long half-life of $^{148}$Gd allowed the transportation of the samples to an underground laboratory in order to reduce the background (see sect.\,\ref{sec:background}).

The $^{7}$Be(p,$\gamma$)$^{8}$B reaction is the starting point of the third pp-chain of hydrogen burning. The $^{8}$B produced $\beta$-decays to $^{8}$Be with a half-life of 770\,$\pm$\,3\,ms. Then, $^{8}$Be promptly decays into two $\alpha$-particles. None of the decays involve $\gamma$-emission and it is preferred experimentally to detect the $^{8}$Be decay $\alpha$-particles over the detection of the $\beta$-decay positrons. Therefore, the $^{7}$Be(p,$\gamma$)$^{8}$B cross section was (with the exception of a recoil mass separator experiment \cite{2000EPJA....7..303G}) always measured by the detection of the $\beta$-delayed $\alpha$-particles \cite[and references therein]{2004PhRvC..70d5801C}. Alpha detection is advantageous in this case also because the $^{7}$Be target is radioactive and emits $\gamma$-radiation, hence the experiment must be performed in a high $\gamma$-background environment caused by the radioactive target.

Because of the 770\,ms lifetime of $^{8}$B, cyclic activation is needed and either the fast transport of the target between the irradiation and the counting position or beam interruptions if the target is not moved.

\subsubsection{X-ray detection}
\label{sec:Xray}

The vast majority of the proton-rich unstable isotopes decay by the competing decay modes of positron emission and electron capture. If the Q value of the $\beta$-decay is less than 1022\,keV, positron emission is forbidden, and thus electron-capture is the only decay mode. An electron, located mainly on the inner (K or L) atomic shells -- whose wave function partially overlaps with the nucleus -- is absorbed by the nucleus and when the remaining vacancy is filled by an outer electron, a characteristic X-ray is emitted. The measurement of the cross section can be based on the detection of the characteristic X-ray radiation. A series of cross section measurements, aiming at the study of the p-process \cite{0034-4885-76-6-066201}, have been carried out at Atomki following this idea. The most important parameters of these measurements are summarized in table \ref{tab:xmeas}.

The characteristic X-ray measurement technique has often clear advantages over the "standard", $\gamma$-detection based activation. Namely, the relative intensity of X-ray radiation is usually high since the electron-capture decay is followed dominantly by the emission of either a K$_{\alpha1}$ or a K$_{\alpha2}$ X-ray (the probability of Auger electron or K$_{\beta}$ X-ray emission is typically (at least) one order of magnitude smaller in the A $\geq$ 100 region). Furthermore, the energies of the emitted characteristic X-rays fall in a narrow region where the efficiency of the widely used germanium detectors is the highest. There are some cases when the reaction product decays without emitting $\gamma$-radiation\footnote{Such reactions, where the $\beta$-decay is followed solely by characteristic X-ray emission are e.g. $^{127}$I($\alpha,\gamma$)$^{131}$Cs or $^{175}$Lu($\alpha,\gamma$)$^{179}$Ta.}, where this technique provides the only opportunity to determine the cross section by activation. 

The disadvantage of this approach is, however, that the characteristic X-ray measurement technique is not able to distinguish between the decay of the different isotopes of the same element. If the target consists of several stable isotopes of the same element, then different isotopes will inevitably be created and the separation of the reactions by X-ray detection is complicated. Therefore, targets having only one single stable isotope are preferred for this technique, or high isotopic enrichment is needed. Even if the target is isotopically pure, it is still possible that, for example, both the ($\alpha,\gamma$) or ($\alpha$,n) reaction channels are open (which is very often the case for heavy p-isotopes in the studied energy ranges). In this case, two isotopes of the same element are produced. To separate the respective cross sections, the source of the emitted characteristic X-rays needs to be identified. This can be done when the half-lives of the produced isotopes are fairly different. In the following the application of this technique will be illustrated by two examples.

\begin{table*}
\center
\caption{\label{tab:xmeas} Some parameters of cross section measurements -- relevant for the astrophysical $\gamma$-process -- carried out by measuring the yield of emitted characteristic X-rays. The target nuclei, the half-lives of the reaction products, the energies of the emitted K$_{\alpha1}$ characteristic X-rays and the experimental technique used to separate the reaction channels are given. The decay data are taken from refs.\,\cite{NDS75,NDS75b,NDS75c,NDS93,NDS107,NDS108,NDS108b,NDS108c,NDS109,NDS110,NDS112,NDS121}}
\begin{tabular}{cccccc}
\hline
\multicolumn{1}{c}{Target} &
\multicolumn{1}{c}{Half-life of the}&
\multicolumn{1}{c}{Half-life of the} &
\multicolumn{1}{c}{Energy of the } &
\multicolumn{1}{c}{Separation} &
\multicolumn{1}{c}{Reference} \\
\multicolumn{1}{c}{nucleus} &
\multicolumn{1}{c}{($\alpha,\gamma$) product} &
\multicolumn{1}{c}{disturbing product} &
\multicolumn{1}{c}{K$_{\alpha1}$ X-ray} &
\multicolumn{1}{c}{technique} &
\multicolumn{1}{c}{} \\
\hline
$^{121}$Sb & 59.41 $\pm$ 0.01 d & 4.1760 $\pm$ 0.0003 d & 27.47 keV & half-life & \cite{PhysRevC.97.045803} \\
$^{127}$I  & 9.689 $\pm$ 0.016 d & 29.21 $\pm$ 0.04 m & 29.78 keV & half-life & \cite{PhysRevC.86.035801} \\
$^{169}$Tm & 500.4 $\pm$ 3.7 d & 6.70 $\pm$ 0.03 d & 52.39 keV & half-life &\cite{Kiss2011419} \\
$^{191}$Ir & 186.01 $\pm$ 0.06 d & 6.1669 $\pm$ 0.0006 d$^*$ & 66.83 keV & half-life &\cite{SZUCS2018396} \\
\hline
$^{115}$In & 1.5913 $\pm$ 0.0092 d & 5.00 $\pm$ 0.02 h & 25.27 keV & decay curve & \cite{PhysRevC.97.055803} \\
$^{195}$Au & 3.0421 $\pm$ 0.0017 d & 1.088 $\pm$ 0.004 d & 70.82 keV & decay curve & \cite{Szucs_Au_to_be_submitted} \\
\hline
\multicolumn{3}{l}{$^*$ \footnotesize{$^{196}$Au, produced via the $^{193}$Ir($\alpha$,n) reaction}}
\end{tabular}
\end{table*}

If the half-life of the ($\alpha,\gamma$) product is about an order of magnitude longer than that of the ($\alpha$,n) product, X-ray counting with adequate timing can be used to identify the origin of the emission as was done in \cite{Kiss2011419,PhysRevC.86.035801,PhysRevC.97.045803,SZUCS2018396}. As an example, the cross section measurement of the $^{169}$Tm($\alpha$,n)$^{172}$Lu and $^{169}$Tm($\alpha$,$\gamma$)$^{173}$Lu reactions is introduced in the following in detail. Because of the large difference in the half-lives of the produced $^{172}$Lu and $^{173}$Lu (table \ref{tab:xmeas}, where  $^{172}$Lu is included as the disturbing reaction product on $^{169}$Tm), it is a reliable assumption that -- even if the reaction cross sections were the same -- the characteristics X-ray yield measured soon after the irradiation is dominated by the decay of $^{172}$Lu. Moreover, in the studied energy range the cross section of the $^{169}$Tm($\alpha$,n)$^{172}$Lu reaction increases rapidly above the reaction threshold and was found to be at least 50 times higher than the cross section of the $^{169}$Tm($\alpha,\gamma$)$^{173}$Lu reaction \cite{KISS201152}. Thus, to determine the cross section of the radiative $\alpha$-capture reaction, the activity measurement had to be repeated later, when the $^{172}$Lu contamination decayed completely out.

In \cite{Kiss2011419} the characteristic X-ray counting was carried out at least three times for each irradiated target.  Firstly, the activities were measured $t_w$ = 1 - 5 hours after the end of the irradiation to determine the $^{169}$Tm($\alpha$,n)$^{172}$Lu cross section and it was assumed that all the measured characteristics X-rays belong to the decay of $^{172}$Lu. Moreover, the $\beta$-decay of the $^{172}$Lu isotope is followed by the emission of several $\gamma$ rays with high relative intensities. By measuring their yield the X-ray counting based $^{169}$Tm($\alpha$,n) reaction cross sections were crosschecked at high irradiation energies. It was found that the results derived by the two different approaches are in agreement within their statistical uncertainties \cite{Kiss2011419}.

\begin{center}
\begin{figure*}
\centering\resizebox{0.8\textwidth}{!}{\rotatebox{0}{\includegraphics[clip=]{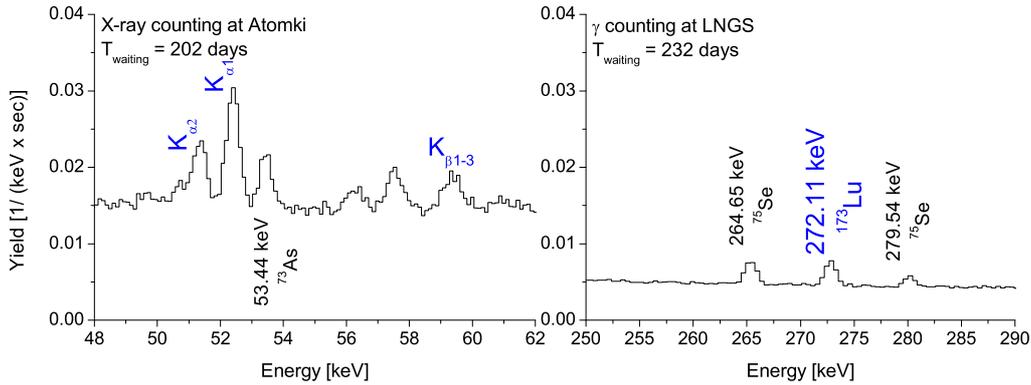}}}
\caption{\label{fig:tm_spectra} Measured characteristic X-ray (left panel) and $\gamma$ spectra (right panel) taken at Atomki and at LNGS, respectively after irradiating a $^{169}$Tm target with an E$_{\alpha}$ = 13.5 MeV $\alpha$ beam. The X-ray and $\gamma$ transitions used for the analysis (blue) and the $\gamma$-lines belonging to parasitic reactions on backing impurities (black), are resulting in the production of $^{73}$As and $^{75}$Se isotopes. It is worth noting that the signal-to-background ratio is similar in the activity measurements carried out at LNGS and Atomki.}
\end{figure*}
\end{center}

After $t_w$ = 168 days and $t_w$ = 200 days from the irradiation the X-ray countings were repeated. During the cooling period of more than 24 weeks, the $^{172}$Lu activity of the targets decreased by a factor of more than 10$^7$, therefore, one can assume that the observed X-ray yield belongs solely to the decay of $^{173}$Lu, produced by the $^{169}$Tm($\alpha$,$\gamma$) reaction. The two different countings resulted in cross sections agreeing within their statistical uncertainties proving that the $^{172}$Lu isotope decayed out completely. Moreover, the electron capture decay of $^{173}$Lu is followed by the emission of the E$_{\gamma}$ = 272.1 keV $\gamma$-line which has a three times lower relative intensity than the K$_{\alpha1}$ transition. The typical laboratory background did not allow using this line to verify the above discussed experimental method. Therefore, two targets (irradiated with E$_{\alpha}$ = 17.5 MeV and E$_{\alpha}$ = 13.5 MeV) were transported to the Laboratori Nazionali del Gran Sasso (LNGS) deep underground laboratory and with a well-shielded 100\% relative efficiency HPGe detector the activity of the targets were measured (see Sec\,\ref{sec:background}). Figure \ref{fig:tm_spectra} shows the X-ray and gamma spectra measured at Atomki and at LNGS. The cross section determined at Atomki by measuring the characteristic X-rays and the ones measured at LNGS were in agreement within statistical uncertainties which proves the applicability of the approach \cite{KISS201152}. 

The method of carrying out two separate countings with long time separation was also used for the measurement of the cross sections of the $^{121}$Sb($\alpha$,$\gamma$), $^{127}$I($\alpha$,$\gamma$/n), $^{191}$Ir($\alpha$,$\gamma$/n), $^{193}$Ir($\alpha$,n) reaction cross sections \cite{PhysRevC.86.035801,PhysRevC.97.045803,SZUCS2018396}. 

If the half-lives of the reaction products are shorter than a few days (but the half-life of the isotope with lower production cross section is at least a few times longer than that of the other  -- disturbing -- isotope of the same element) it is technically feasible to measure the decay curve long enough to fit the sum of two exponential functions characterizing the decay processes. At the beginning of the counting the measured characteristic X-ray yield is dominated by the decay of the shorter-lived isotope. However, after sufficient waiting its activity decreases to a level low enough and the contribution to the total X-ray yield of the longer half-life isotope becomes dominant. As an example, in the following the cross section measurement of the $^{115}$In($\alpha$,$\gamma$)$^{119}$Sb and $^{115}$In($\alpha$,n)$^{118}$Sb$^m$ reactions is discussed in some detail.

\begin{figure}
\includegraphics[width=\columnwidth]{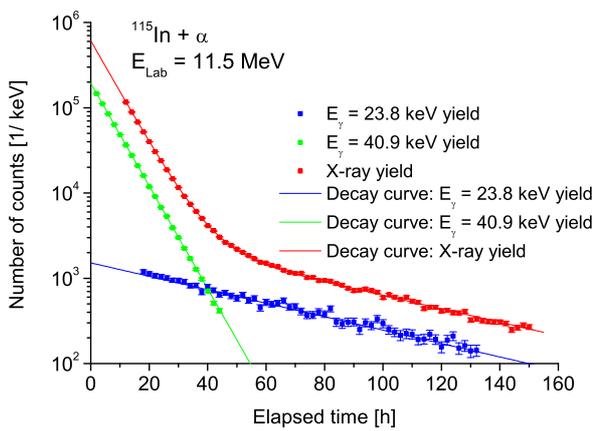}%
\caption{\label{fig:in_decay} Decay curves of the transitions used to determine the $^{115}$In($\alpha,\gamma$)$^{119}$Sb and $^{115}$In($\alpha$,n)$^{118}$Sb$^m$ reaction cross sections. The measured characteristic X-ray yields were fitted with the sum of two exponential functions of known half-lives. The shift between the curves is caused by the different relative intensities of the X-rays and $\gamma$-rays.}
\end{figure}

The element indium has two stable isotopes $^{113}$In and $^{115}$In with natural abundances of 4.29\,$\pm$\,0.04\,\% and 95.71\,$\pm$\,0.04\,\%, respectively. Alpha-induced reactions lead to unstable Sb isotopes, with half-lives of 15.8\,$\pm$\,0.8 min ($^{116}$Sb$^g$), 60.3\,$\pm$\,0.6 min ($^{116}$Sb$^m$), 2.80\,$\pm$\,0.01 h ($^{117}$Sb), 3.6\,$\pm$\,0.1 min ($^{118}$Sb$^g$), 5.00\,$\pm$\,0.02 h ($^{118}$Sb$^m$) and 38.19\,$\pm$\,0.22 h ($^{119}$Sb). Although several reaction channels are energetically accessible and thus several different unstable Sb isotope were produced, the half-life of the $^{119}$Sb isotope is far the longest, therefore, by measuring long enough the yield of the emitted X-rays, the radiative $\alpha$-capture cross section of $^{115}$In could be derived. Figure \ref{fig:in_decay} shows the yield of the time-dependent characteristic X-rays together with the fitted exponential functions. As can be seen, in the first dozen spectra the characteristic X-ray yield is dominated by the decay of the $^{118}$Sb$^m$ nuclei, however, after about 48-50 hours, the contribution from the $^{119}$Sb decay starts to prevail. Moreover, the decay of both the $^{118}$Sb$^m$ and $^{119}$Sb isotopes is followed by the emission of $\gamma$-rays. The cross sections were measured by counting the 23.9 keV ($^{119}$Sb) and 40.8 keV ($^{118}$Sb)$\gamma$-rays, as well. The agreement between the alpha-induced cross sections based on X-ray and $\gamma$ countings were found to agree within 4\%, better than the independent uncertainties of the two values. A similar experimental approach to the one described here was used to measure the $^{197}$Au($\alpha$,$\gamma$/n) cross sections \cite{Szucs_Au_to_be_submitted}. 

For the technical aspects of X-ray detection, LEPS (Low Energy Photon Spectrometer) or LEGe (Low Energy Germanium detector) detectors are the most suitable HPGe detectors for such measurements. The cylindrical Germanium crystal of these detectors is relatively thin but the surface is large and the entrance window of the detector is thin. Accordingly, the detection efficiency for the K$_{\alpha}$ and K$_{\beta}$ lines is high but the detector is largely insensitive to the $\gamma$-rays in the MeV energy range owing to the small thickness of the crystal. Further, the resolution of these detectors is outstanding (e.g. for a GL2015R type LEPS used in the experiments performed at Atomki, the FWHM of the 5.9 keV line of $^{55}$Fe source is 380 eV and the FWHM of the 122 keV line of $^{57}$Co source is 600 eV), enabling the separation of the K$_{\alpha1}$ and K$_{\alpha2}$ lines for an A $\geq$140 reaction product - as it is shown in fig. \ref{fig:tm_spectra}. It is worth noting that for measuring low or moderate $\gamma$-ray energies, LEPS detectors may well outperform the much more expensive large volume coaxial detectors \cite{doi:10.1063/1.4875306,PhysRevC.94.045801}.

\subsubsection{Detection of annihilation radiation}
\label{sec:annihilation}

Stellar hydrogen burning processes typically involve proton capture reactions. If a radioactive species is produced by such a reaction, it is located on the proton-rich side of the valley of stability and decays by positron emission or electron capture. In the lower mass region, where the hydrogen burning processes are important, positron decay is dominant (with just a few exceptions, like $^{7}$Be, where energetically only the electron capture is allowed). Often positron emission leads to the ground state of the daughter nucleus with no $\gamma$-radiation involved. An activation experiment in such a case can be performed by the detection of the positron annihilation radiation.

In a metallic environment of a typical target backing the positrons emitted by the decay annihilate directly without the formation of an intermediate positronium \cite{humberston2012atomic}. The annihilation takes place predominantly by the creation of two 511\,keV energy $\gamma$-quanta which are emitted in opposite directions and are well suited for the reaction cross section determination. Reactions of astrophysical interest which can be investigated with this method are, among others, $^{12}$C(p,$\gamma$)$^{13}$N \cite{PhysRev.77.194,PhysRev.77.197}, $^{14}$N(p,$\gamma$)$^{15}$O \cite{doi:10.7566/JPSCP.14.020403}, $^{14}$N($\alpha,\gamma$)$^{18}$F \cite{PhysRevC.62.055801}, $^{17}$O(p,$\gamma$)$^{18}$F \cite{PhysRevLett.95.031101,PhysRevC.75.035810,PhysRevLett.109.202501,PhysRevC.89.015803,PhysRevC.95.035805}.

In activations based on the detection of annihilation radiation, the identification of the reaction product is especially crucial. This follows from the fact that the 511\,keV radiation may be due to different sources. It is always present in the laboratory background, but if it is guaranteed that the background does not change in time, this background can easily be subtracted. More problematic are additional reactions on target impurities which produce positron emitting radioactive species. The pair production of high energy $\gamma$-rays also leads to 511\,keV $\gamma$'s. It is especially important therefore to avoid and control target impurities and test runs with blank targets may also be necessary. 

\begin{figure}
	\includegraphics[angle=270,width=\columnwidth]{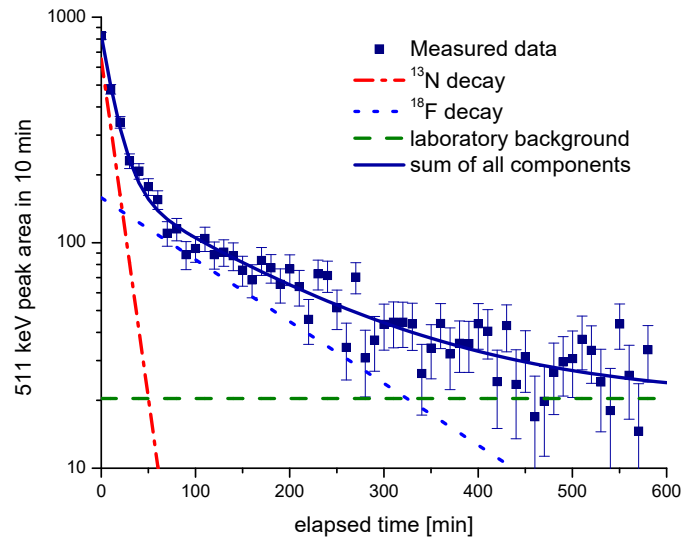}%
	\caption{511\,keV peak area as a function of time measured after the proton irradiation of a $^{17}$O target with a $^{12}$C impurity. Three important sources of 511\,keV radiation are identified.}
	\label{fig:annih_decay}
\end{figure}

Even with such precautions, it is of utmost importance to follow the decay and analyze the half-life of the reaction product in order to identify any possible contamination. Figure\,\ref{fig:annih_decay} shows an example of such an analysis. This decay curve was measured during the study of the $^{17}$O(p,$\gamma$)$^{18}$F reaction at an energy which is close to the resonance of the $^{12}$C(p,$\gamma$)$^{13}$N reaction at 457\,keV. $^{18}$F has a half-life of 110\,min while the half-life of $^{13}$N is 10\,min. At this particular energy, the $^{12}$C(p,$\gamma$)$^{13}$N cross section is high enough that the small carbon build-up on the target results in a strong $^{13}$N activity. As is seen in the figure, the decay of the 511\,keV signal can be well fitted by three components: $^{18}$F decay, $^{13}$N decay and laboratory background. With such an analysis, the $^{17}$O(p,$\gamma$)$^{18}$F cross section could be obtained in a reliable way \cite{PhysRevC.95.035805}. 

A precise efficiency determination of the $\gamma$-detector in the case of the 511\,keV $\gamma$-detection requires special attention. If the target is transported from the irradiation chamber to an off-line counting setup, the target must be placed in front of the detector in such a way that a point-like source is obtained. This means that the target must be surrounded by thick enough material to stop the created positrons completely. This can be achieved typically by less than 1\,mm of a heavy metal which does not attenuate too much the 511\,keV $\gamma$-rays. In the case of short half-lives (such as the $^{15}$O (t$_{1/2}$\,=\,2.023\,$\pm$\,0.003\,min) produced in the $^{14}$N(p,$\gamma$)$^{15}$O reaction), the decay counting may be carried out directly at the activation chamber. In this case the positrons emitted towards the vacuum can leave the target and the annihilation takes place in an extended geometry. In such a situation the efficiency determination requires dedicated experiments aided preferably by simulations. 

\subsubsection{Electron detection \label{sec:electron_detection}}

\begin{figure}[ht]
 \includegraphics[width=\columnwidth]{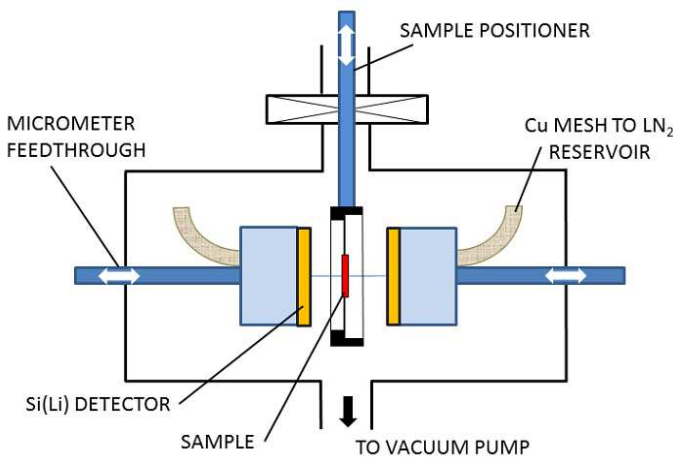}
 \caption{\label{4piespec} Schematic setup of the 4$\pi$ 
	electron spectrometer. (Figure adapted from Ref. \cite{RAK04}). }
\end{figure}

The neutron magic nuclei $^{208}$Pb and $^{209}$Bi constitute the 
termination point of the s-process reaction chain. Activation measurements
for these isotopes are difficult because the induced activities exhibit no or 
very weak $\gamma$ transitions, respectively. Therefore, a 4$\pi$ electron 
spectrometer consisting of two Si(Li) detectors in close geometry was used 
to measure the $\beta$ spectra emitted from the reaction products 
\cite{RAK04}. fig.~\ref{4piespec} shows a sketch of the setup. The Si(Li) 
detectors were mounted on calibrated micrometers so that their distance
during the measurement could be adjusted to less than 1 mm, resulting 
in an effective solid angle close to 4$\pi$. 

As the detectors were kept at a temperature of 110 K, a sluice allowed 
the samples to be changed without breaking the vacuum. In the relevant 
electron energy interval between 15 and 1000 keV, the general room background 
could be reduced to a rate of 0.49 s$^{-1}$. Electron backscattering effects
were minimized by summing the signals of both detectors. The high efficiency 
of the spectrometer of $93.5\pm1.6$\%, which was determined by comparing
the $\beta$- and $\gamma$-activities of activated gold foils, was essential
for obtaining the extremely small MACS of the doubly neutron magic 
nucleus $^{208}$Pb with an uncertainty of 5\,\%, which is essentially determined by the flux measurement and the $\beta^-$-efficiency.

\begin{figure}[ht]
 \includegraphics[angle=270,width=0.9\columnwidth]{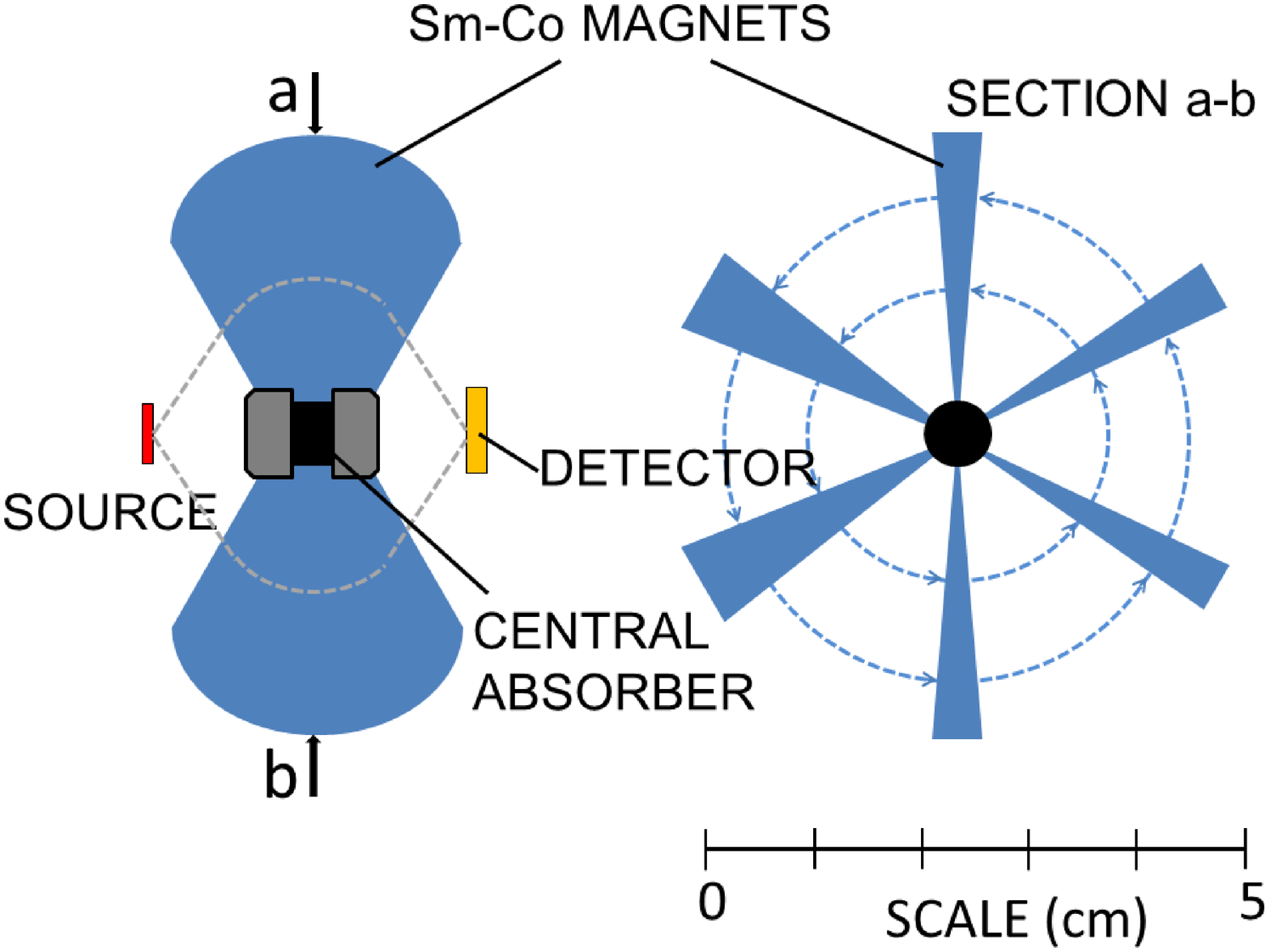}
 \caption{\label{miniorange} Schematics of a mini-orange spectrometer. (Figure adapted from Ref. \cite{WWK82}). }
\end{figure}

In measurements of radioactive samples the electron component of the 
induced activity is obscured by intense backgrounds of $\gamma$ rays, 
$\alpha$ particles, or X rays. A mini-orange spectrometer provides a 
solution for the selective detection of electrons \cite{VaW72,VFG78}.
This compact version of an orange-type beta spectrometer uses
small permanent magnets for the production of a toroidal magnetic 
field as sketched in fig. \ref{miniorange}. The six wedge-shaped 
samarium cobalt permanent magnets are magnetized perpendicular 
to their largest surface and produce a toroidal field in the gaps to bend
electrons emitted from the sample around a central absorber towards 
a Si(Li) detector, while background radiation from the sample is blocked 
by the central tungsten absorber. The electron transmission curve  
can be tailored by adapting the shape of the permanent magnets so
that a large part of the investigated beta spectrum can be observed
in a single measurement. 

A first example for the successful use of mini-orange spectrometers was
the measurement of the isomeric ratio of $^{241}$Am \cite{WWK82},
which determines the relative production of $^{242}$Cm and $^{244}$Cm 
in nuclear reactors, where the low-energy conversion electrons from the 
decay of $^{241}$Am could be strongly suppressed. A second application 
was chosen for detecting the weak electron branch in the decay of the 3.9 
min isomer in $^{79}$Se that allowed to quantify the temperature-dependence 
of $^{79}$Se under s-process conditions \cite{KlK88}.

\subsection{A note on the importance of decay parameters}
\label{sec:decaypar}

The cross section determination based on an activity measurement requires knowledge of the decay parameters of the produced isotopes. This is in contrast to the technique of Accelerator Mass Spectroscopy (AMS, see sec. \ref{sec:AMS}), which - due to direct atom counting - is independent of any decay characteristics. The decay parameters are the half-lives, the energies and relative intensities of the decay radiations. If the isotopes of interest are only a few mass units away from stable isotopes, they have reasonably long half-lives, thus they could be studied extensively in the last few decades. Therefore, these decay parameters are usually relatively well known. In some cases, however, less precise or apparently wrong parameters can also be encountered, especially on the neutron-deficient side of the chart of nuclides. In these cases an additional systematic uncertainty has to be added for the cross section determination, or a dedicated experiment must be performed for the precise measurement of the decay parameter in question.

The energy of the decay radiation is usually the least crucial parameter. In the case of $\gamma$-counting the energy of the emitted $\gamma$'s is usually well known from decay experiments or from the known level schemes of the daughter nuclei. In addition, the energy information is not necessary for the cross section determination, it is needed only for the identification of the $\gamma$-ray to be measured, which usually does not pose a severe difficulty.

\begin{figure}
	\includegraphics[angle=270,width=\columnwidth]{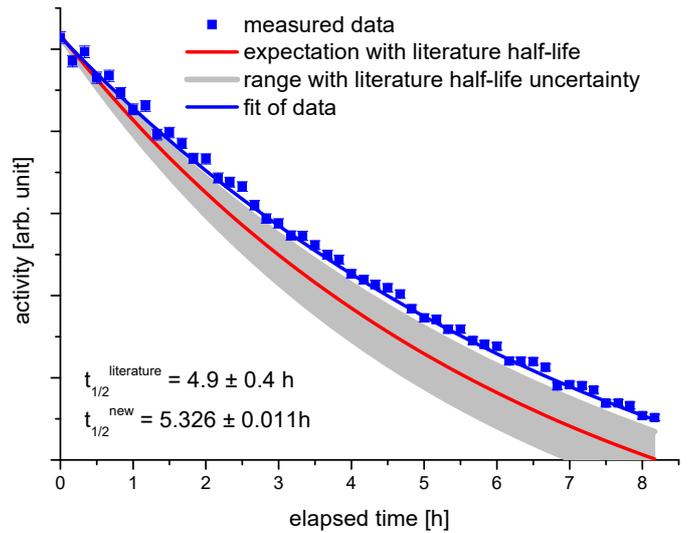}%
	\caption{Decay of $^{133}$Ce. The red solid line shows the expected decay based on the literature half-life with the uncertainty indicated by the gray band. The measured data clearly deviate from the expectation. Based on the measurement a new, more precise half-life could be determined.}
	\label{fig:133Ce_decay}
\end{figure}

The half-lives and relative intensities of decay radiation enter into the cross section calculations and their uncertainty impacts on the final accuracy of the cross section. Half-life values with uncertainties of more than a few percent can be encountered. Often the measurement of the decay of the reaction products can reveal uncertain half-lives. An example is given in fig.\,\ref{fig:133Ce_decay} for the decay of $^{133}$Ce, produced by the $^{130}$Ba($\alpha$,n)$^{133}$Ce reaction \cite{PhysRevC.85.025804}. The measured data clearly indicate a deviation from the literature half-life and the measurement allowed a much more precise half-life determination.

In several cases, an activation cross section measurement leads also to half-life determination as a useful side result. Improved precision half-life values were recently provided e.g. for the following isotopes: $^{109}$In \cite{PhysRevC.71.057302}, $^{110}$Sn \cite{PhysRevC.71.057302}, $^{133}$Ce \cite{Farkas2011}, $^{154}$Tb \cite{2009NuPhA.828....1G} and the analysis on further isotopes ($^{65}$Ga, $^{95}$Ru, $^{95}$Tc, $^{125}$Cs, $^{125}$Xe, $^{144}$Pm) is ongoing. The photoactivation technique (sec.\,\ref{sec:gammainduced}) was used to measure the half-life of several Au, Hg and Pb isotopes \cite{PhysRevC.63.047307}.

Appropriate knowledge of the relative intensity of the decay radiations is more difficult to obtain. Their measurement requires the determination of the source activity independently and may involve therefore an additional measurement on its own. For cross section determination, the relative intensities are thus usually taken from the literature. 

In some cases the uncertainty of the relative intensities is one of the largest systematic uncertainties of the measured cross section. Just to give one example: The cross sections of the $^{64}$Zn(p,$\gamma$)$^{65}$Ga and $^{64}$Zn(p,$\alpha$)$^{61}$Cu reactions were measured recently by activation \cite{PhysRevC.90.052801}. For both cases, $^{65}$Ga and $^{61}$Cu, the relative intensities of the individual $\gamma$-transitions used for the analysis are known well, but their absolute normalization is highly uncertain by almost 20\,\% \cite{BHAT1999417,BROWNE20102425} owing to the difficulty of measuring absolute $\gamma$-intensities. As the uncertainty of the intensities enters linearly into the cross section calculation, new determination of these absolute intensities would be needed in order to reduce the uncertainty of the measured cross sections.

\subsection{Background considerations}
\label{sec:background}

The background issues of activation experiments were mentioned several times in the previous sections. In general, background causes less problem in activation compared to in-beam experiments. In nuclear astrophysics, in-beam experiments often deal with low cross sections which are to be measured by particle or $\gamma$-detection during the bombardment of the target by the beam. Any target impurity, especially light isotopes having high cross sections may cause prompt radiation which can disturb the detection of the searched for signal. In the case of $\gamma$-detection, reactions producing high energy $\gamma$-rays are especially problematic as they cause Compton background which may cover the weak signals of the studied reaction. Such disturbing reactions include  $^{11}$B(p,$\gamma$)$^{12}$C or $^{19}$F(p,$\alpha\gamma$)$^{16}$O in the case of proton induced reactions, or $^{13}$C($\alpha$,n)$^{16}$O in the case of alpha induced reactions. In the latter case the capture of the produced neutron causes $\gamma$-background.

In activation experiments, the prompt radiations during the beam bombardment are not relevant. Conversely, the activation of nuclides other than the one of interest needs to be monitored especially for those having half-lives comparable to or longer than the one of the investigated isotopes. Common choices for target backings are aluminum or some heavy element like tantalum. 

It is important to use high-purity materials both for target and backing. Even impurities of very low concentration may cause severe background problems if they exhibit high cross sections. A common impurity in targets is iron. Different radioactive Co or Ni isotopes are produced on stable Fe isotopes by proton or $\alpha$-bombardment which can be disturbing low activity measurements. 

Elemental targets with more than one stable isotope sometimes suffer from a dominant reaction on one of the isotopes with a much higher cross section than the reaction under investigation; or the decay of many isotopes may produce too high a background. In such cases, isotopically highly-enriched target materials are necessary. Enrichment is also necessary if the natural abundance of the studied isotope is too low. This is often the case for $\gamma$-process experiments (where proton rich heavy isotopes with typically very low natural abundances are studied) or for $^{3}$He($\alpha,\gamma$)$^{7}$Be where $^{3}$He has a natural abundance of only 0.000134\,\%.

If beam-induced activities can be avoided, there may still be a problem with the ambient laboratory background. In an activation experiment the detector can be shielded in a 4$\pi$ geometry which reduces strongly the laboratory background. Cosmic ray induced interactions, however, cannot be avoided by such a passive shielding. Active methods, like the application of cosmic-ray veto detectors can be useful (see e.g. \cite{2002NuPhA.710..469O,a4f5e5a33acd43bfa952482c6f955508}). If the half-life of the reaction product is long enough, the irradiated target can be transported to e.g. an underground counting facility which provides ultra low background conditions \cite{Best2016}. For instance,  samples for studying the $^{3}$He($\alpha,\gamma$)$^{7}$Be \cite{PhysRevLett.97.122502,PhysRevC.75.035805}, the $^{144}$Sm($\alpha,\gamma$)$^{148}$Gd \cite{1998A&A.333.1112S} and the $^{169}$Tm($\alpha,\gamma$)$^{173}$Lu \cite{KISS201152} reactions were counted in the low background facility \cite{LAUBENSTEIN2004167,LNGSlowbck} of the LNGS deep-underground laboratory in Italy \cite{LNGS}.

\section{Accelerator Mass Spectrometry}
\label{sec:AMS}

\subsection{Introduction}

AMS is a single-atom counting technique, usually used to measure minute amounts of nuclides in mg-sized samples. AMS represents the most sensitive technique for quantifying long-lived radionuclides \cite{Syn13,Kut13,Kut16}. It is a mass spectrometric technique based on the use of a (tandem) accelerator. This method completely suppresses any molecular isobaric (same mass) interferences. With the high particle energies provided by the accelerator it is capable of separating atomic isobars using dedicated particle detectors.

For specific reactions, AMS offers a powerful tool to measure cross sections. The advantage compared to decay counting is that it is independent of the half-lives of the reaction products. The combination of activation and subsequent AMS measurement was applied for a range of measurements where off-line decay counting is difficult or impossible due to long half-lives of reaction products or due to weak or missing $\gamma$-ray transitions. The interested reader will find a comprehensive summary of AMS in nuclear physics and astrophysics up to 1990 in \cite{KP90} and in the subsequent AMS proceedings \cite{AMS1_2,AMS_NIM}. AMS was also used for measuring half-life values of long-lived radionuclides \cite{Kut13,Kut16,KP90,KPK12}; and also for the search of superheavy elements \cite{LFK12,DWF11,DFR11,KK15} and for other exotic rare nuclides \cite{Kut13,Kut16,KP90}.

AMS's advantage compared to conventional (low-energy) mass spectrometers is its unsurpassed sensitivity for the abundance of specific nuclides. The typical applications deal with abundance ratios (radionuclide or rare isotope versus stable isotope) in samples that differ by 12 to 17 orders of magnitude, and is usually applied to the detection of radionuclides with half-lives ranging from years up to a hundred million years. Such abundance sensitivities allow the detection of these isotopes at natural concentrations with applications in radiocarbon ($^{14}$C) dating, climate research, from environmental research to biomedical applications and nuclear (astro)physics. Overall, about 30 radionuclides are used in present AMS applications.

Decay counting becomes difficult in case of long-lived reaction products. However, the technique of AMS does not measure decays, but directly the number of produced radionuclides. Direct atom counting of the reaction products via AMS provides therefore a powerful complement of the conventional activation method as it is essentially independent of the half-life and decay characteristics of the reaction product, thus reducing the related uncertainties of the traditional activation technique. AMS, as any activation method (see section \ref{sec:actmethod}), inherently includes the direct capture contribution. 

AMS is used to measure cross sections of nuclear reactions that often cannot be measured using conventional decay-counting approaches. It is best suited to reactions where the product nucleus has a long half-life and/or missing or uncertain $\gamma$-ray transitions which restrict the measurement of the decay activity. This method can provide accurate anchor points for critical reactions leading to radioactive nuclides for specific energies with the important aspect of being fully complementary and independent to previous experiments. 

AMS was introduced to laboratory experiments in nuclear astrophysics already in 1980 just at the advent of the AMS technique itself by Paul {\it et al.} \cite{PHK80} 
with a first study of the $^{26}$Mg($p, n$)$^{26}$Al reaction. However, only in the last 10 years this technique had been applied more routinely for cross section measurements of charged particle and neutron-induced reactions. The list of reaction products of interest for astrophysics includes radioisotopes over the entire mass range, e.g. $^{10}$Be, $^{14}$C, $^{26}$Al, $^{36}$Cl, $^{41}$Ca, $^{44}$Ti, $^{55,60}$Fe, $^{59,63}$Ni, $^{68}$Ge, $^{79}$Se, $^{129}$I, $^{182}$Hf (see e.g. \cite{PFA03,NPA06,AFF06,NPA05a,WBD07,DDH09,Wal10,DFK10,RDF07,LRD16,WBB13,RGC12,ASC17,WBB17,VBW08,WBB16}). Furthermore, neutron capture on $^{209}$Bi and actinides ($^{232}$Th, $^{235,238}$U) were studied as well as a number of actinide isotopes including interstellar $^{244}$Pu (see e.g. \cite{Pau01,Wal04,WFF15,WBB14}) and search was carried out for existence of superheavies (\cite{LFK12,DWF11,DFR11,KK15}). Many of the neutron-induced reactions utilized a spectrum imitating almost perfectly a Maxwell-Boltzmann distribution for kT = 25 keV (see section \ref{sec:neutroninduced.2}) \cite{RaK88}. AMS laboratories with an active program for cross section measurements within the last two decades include TU Munich \cite{DFK10,RDF07,LRD16}, ETH Zurich \cite{Syn13}, Caserta \cite{LSF10}, ANL \cite{NPA06,KPK12,PHK80,K12,K08}, Purdue \cite{ASC17}, Notre Dame \cite{RGC12,BKB13,Bow13}, the ANU \cite{WBB16,WBB17,PCF18}, CIAE \cite{DZX15} and the VERA facility \cite{WBB16,Kut13,Kut16,Wal10,WBB17} at the Univ. of Vienna (more details are given e.g. in Tab. \ref{tab:xs_AMS_list}).
Many more reactions are studied that are relevant to nuclear technology (nuclear fusion, nuclear fission and advanced reactor concepts, but also for medical applications and radiation dose estimations). Additional measurements include environmental, geological and extraterrestrial applications that rely on production rates of cosmogenic radionuclides.

\subsection{AMS and cross-section measurements}

We note that AMS cannot be considered a general technique for all reaction products because of the limited number of radionuclides developed so far. The method is a 2-step process: the first step is irradiation of a sample (based on the standard activation technique). The second step is the subsequent AMS detection of the reaction product; i.e. the number of product (long-lived) radionuclei is quantified by mass spectrometry.
Rewriting eq. \ref{eq:crosssec} (see section \ref{sec:actmethod}), we obtain

\begin{equation}
\label{eq:sig}
\sigma_{\rm exp} = \frac{\rm N_{prod}}{\rm N_{target}}\times\frac{1}{\Theta_{\rm tot}}
\end{equation}
where $\rm N_{prod}$ is the quantity given in eq.\,\ref{eq:production1}.

The experimental cross section can simply be calculated by two quantities, the isotope ratio N$_{prod}$/N$_{target}$  (conversion ratio), which is directly measured by AMS, and the fluence $\Theta_{\rm tot}$ (i.e. the time-integrated fluence rate [particles/cm$^2$]). The fluence is usually determined independently, e.g. in case of neutron irradiation from gold monitor foils simultaneously irradiated with the samples. Note the particular advantage of the AMS method, i.e. that the cross section is determined by the measured isotope ratio only, completely independent of the sample mass and the decay properties of the product nucleus.

The reaction product is usually counted relative to a stable isotope of the same element. In case of neutron-capture studies this allows analyzing directly the irradiated sample material. Chemical sample preparation is, however, required in cases where the reaction product is of a different element. In these cases, following activation, the desired radionuclides will need to be separated from the bulk material. A stable tracer material of the same element as the reaction product is added and the radionuclide is measured relative to this tracer. 

To illustrate the high sensitivity of AMS as an atom-counting technique, the cases of $^{13}$C($n, \gamma$)$^{14}$C and \linebreak $^{64}$Zn($\alpha, \gamma$)$^{68}$Ge are selected, leading to the two radionuclides $^{14}$C (t{$_{1/2}$}=5730 yr) and $^{68}$Ge (t{$_{1/2}$}=271 days), respectively. The machine background for AMS measurements of $^{68}$Ge and $^{14}$C is of the order of  $^{68}$Ge/Ge and $^{14}$C/$^{12}$C = $10^{-16}$ atom/atom. This number is obtained from measurements of blank samples which are assumed to contain negligible amounts of the respective radionuclide (Tab. \ref{tab:AMS_RN}). The fraction of particles finally detected in AMS with a particle detector, is a few percent for $^{14}$C and $^{68}$Ge. 

AMS uncertainties are rarely better than 3\,\%, independent of counting statistics. If we require a similar uncertainty in the counting statistics, a 1\,\% overall efficiency in AMS requires to sputter $10^{5}$ radionuclides. This is the minimum number of atoms that need to be produced in the sample activation. This minimum number corresponds to activities between $\mu$Bq and mBq for $^{14}$C and $^{68}$Ge, respectively. In neutron irradiations, the typical neutron fluence rates are between $10^{9}$ and $10^{10}$ neutrons cm$^{-2}$ s$^{-1}$ [see e.g. Karlsruhe Institute of Technology (KIT) or Soreq Applied Research Accelerator Facility (SARAF)]. With reasonable activation times of days to one week, a total fluence of $10^{14}$ to $10^{15}$ neutrons cm$^{-2}$ can be achieved.
In case of charged-particle induced reactions, e.g. Atomki \cite{PhysRevC.74.025805} produces $\mu$A $\alpha$-beams, a 2 days irradiation time results in a total dose of 1$\times$$10^{18}$ $\alpha$ particles. Under these conditions cross sections well below the $\mu$barn level can in principle be studied.

Using AMS, sample masses of order 50 mg are sufficient for determining the conversion ratio in a nuclear reaction. This ratio is only a function of the particle (neutron) fluence and the reaction cross section (Eq. \ref{eq:sig}), independent of the sample mass and, due to the low masses used, not affected by multiple scattering corrections.

\subsection{AMS technique}

In most cases (due to the use of tandem accelerators), negatively-charged ions are produced in a Cs-sputter source, usually from sputtering solid samples. The typical sample masses are of the order of mg per sputter sample. The sample material itself is used up during the measurement. The negative ions are pre-accelerated, energy- and mass-selected by passing electrostatic deflector and/or an injection magnet, respectively. The ions are injected into a tandem accelerator. At the terminal of the accelerator a gas or foil stripper is utilized to strip-off electrons. The ions then leave the accelerator positively charged after being accelerated a second time. 

The acceleration and the stripping process leads to the destruction of molecular isobars. A second analyzing magnet is used to select a specific positive charge state and all molecular break-up products are deflected. AMS completely destroys and removes molecular ions in the beam with the use of the accelerator and a subsequent second mass filter (analysing magnet). Stable isobars, however, follow the same path through the accelerator and subsequent analysers, and can be present at levels many orders of magnitudes (up to 10$^{10}$) more abundant than the rare isotope. As a consequence, atomic isobars cannot be removed from the beam by selective filtering.

Several approaches can be made to reduce isobaric interference, e.g.: specialised sample preparation can reduce the amount of the stable isobar; selecting either an elemental ion or a particular molecular ion can reduce the intensity of the isobar significantly; or spatial separation of isobars can be achieved in a gas-filled magnetic spectrograph \cite{PGH89,Syn13,Kut13,Kut16}. By taking advantage of the different energy loss behaviour of different elements in dedicated particle detectors allows them to be distinguished in principle, AS the ions are identified by their position, energy, energy loss signals and their entering angle. Other isobar separation techniques include selective photo-detachment using lasers, x-ray detection or a particle detection setup consisting of a passive absorber and a time-of-flight system \cite{Syn13,Kut13,Kut16}. 

At best (e.g. $^{14}$C, $^{55}$Fe \cite{WBB16,WBB17}), a measurement reproducibility of 1\,\% can be achieved in AMS,  although more typically, the final AMS uncertainties are of order 3--5\%. Radionuclides with a strong isobaric interference require large particle accelerators that provide the high particle energies of order 100\,MeV to 200\,MeV (this is several MeV/amu) resulting in a better discrimination of radionuclide against interfering isobar in the particle detector. Measurements at such facilities will have larger uncertainties between typically 7 and 10\,\% or higher (see also Tab. \ref{tab:AMS_RN}). In this regard, an important aspect is the availability of accurate AMS standards which are required for absolute normalization. Possible long-term drifts of the particle transmission along the beamline have to be monitored. Therefore, for quality control, the transmission is regularly monitored by means of standards with well-known isotope ratios. Because inherent effects such as mass fractionation, machine instabilities, or potential beam losses between the current measurement and the respective particle detector are difficult to quantify in an absolute way to better than 5 to 10\%, accurate AMS measurements depend on well-defined reference materials. Therefore, together with samples of unknown isotope ratios, reference materials with well-known ratios are measured in periodic intervals as well.

\begin{table*}[bt]
\caption{List of some radionuclides measured by AMS.
\label{tab:AMS_RN}}
\begin{tabular} {lrrrrl}
\hline\hline
Radionuclide		& Half-life	    	& AMS overall  			& Detection limit$^a$ 		& Precision$^a$ 	& Remark	\\
				& 			& efficiency 			&  \\
\hline
$^{10}$Be 			& 1.39 Myr		& 0.1 \% 			&  $\sim$10$^{-16}$			& 2 \%			& stable isobar: $^{10}$B  		 \\
$^{14}$C			 	& 5730 yr		& few \% 			&  \textless10$^{-16}$		& 0.15 \%			& no isobaric interf.  ($^{14}$N$^{-}$ not stable)  		 \\
$^{26}$Al			 	& 0.72 Myr		& 0.02 \% 			&  $\sim$10$^{-16}$			& 2--3 \%			& no isobaric interf. ($^{26}$Mg$^{-}$ not stable) 		 \\
$^{32}$Si		 		& 153 yr		& 0.1 \% 			&  $\sim$10$^{-15}$			& 5 \%			& stable isobar: $^{32}$S  	 \\
$^{36}$Cl			 	& 301 kyr		& few \% 			&  $\sim$10$^{-16}$			& 2--3 \%			& stable isobar: $^{36}$S  	 \\
$^{39}$Ar			 	& 269 yr		& \textless 0.1 \% 	&  $\sim$10$^{-16}$			& 10 \%			& stable isobar: $^{39}$K  		 \\
$^{41}$Ca	 		& 103 kyr		& 0.1 \% 			&  $\sim$10$^{-15}$			& 3 \%			& stable isobar: $^{36}$S  	 \\
$^{53}$Mn		 	& 3.7 Myr		& 0.01 \% 			&  $\sim$10$^{-14}$			& 5 \%			& stable isobar: $^{53}$Cr 	 \\
$^{55}$Fe			 	& 2.7 yr		& 0.1 \% 			&  $\sim$10$^{-16}$			& 2 \%			& no isobaric interf. ($^{55}$Mn$^{-}$ not stable)   	 \\
$^{60}$Fe			 	& 2.6 Myr		& 0.05 \% 			&  $\sim$10$^{-17}$			& 3--5 \%			& stable isobar: $^{60}$Ni  	 \\
$^{59}$Ni			 	& 75 kyr		& \textless0.01 \% 	&  $\sim$10$^{-13}$			& 7 \%			& stable isobar: $^{59}$Co  	 \\
$^{63}$Ni			 	& 100 yr		& \textless0.01 \% 	&  $\sim$10$^{-13}$			& 10 \%			& stable isobar: $^{63}$Cu  	 \\
$^{68}$Ge		 	& 0.74 yr		&few \% 			&  $\sim$10$^{-16}$			& 5 \%			& no isobaric interf. ($^{68}$Zn$^{-}$ not stable)   	 \\
$^{79}$Se		 	& 327 kyr		& \textless0.01 \% 	&  $\sim$10$^{-12}$			& 10 \%			& stable isobar: $^{79}$Br  	 \\
$^{81}$Kr			 	& 230 kyr		& \textless0.01 \% 	&  $\sim$10$^{-13}$			& 10 \%			& stable isobar: $^{81}$SBr 	 \\
$^{93}$Zr			 	& 1.64 Myr		& \textless0.01 \% 	&  $\sim$10$^{-12}$			& 7 \%			& stable isobar: $^{93}$Nb  	 \\
$^{99}$Tc			 	& 210 kyr		& 	 			&  \textendash				& 7 \%			& no isobaric interference   	 \\
$^{129}$I			 	& 16.1 Myr		& few \% 			&  \textless 10$^{-15}$		& 2 \%			& no isobaric interf. ($^{129}$Xe$^{-}$ not stable)  	 \\
$^{146}$Sm		 	& 68 Myr		& \textless0.01 \% 	&  $\sim$10$^{-12}$			& 5 \%			& stable isobar: $^{146}$Nd  	 \\
$^{182}$Hf		 	& 9.0 Myr		& 0.1 \% 			&  $\sim$10$^{-11}$			& 7 \%			& stable isobar: $^{182}$W  	 \\
$^{202m}$Pb		 	& 52.5 kyr		& \textless0.01 \% 	&  $\sim$10$^{-14}$			& 7 \%			& no isobaric interf. ($^{202}$Hg$^{-}$ not stable)  	 \\
$^{210}$Pb		 	& 22.3 yr		& \textless0.1 \% 	&  $\sim$10$^{-13}$			& 4 \%			& no isobaric interference   	 \\
$^{210m}$Bi		 	& 3.0 Myr		& \textless0.1 \% 	&  $\sim$10$^{-13}$			& 5 \%			& no isobaric interference   	 \\
$^{226}$Ra		 	& 1.6 kyr		&  				&  \textendash				& 5 \%			& no isobaric interference   	 \\
$^{229}$Th		 	& 7.93 kyr		& 	&  \textendash				& 5 \%			& no isobaric interference   	 \\
$^{231}$Pa		 	& 32.8 kyr		& 		&  \textendash				& 4 \%			& no isobaric interference   	 \\
$^{233,236}$U		 	& 0.16 \& 23.4 Myr	& 	&  $\sim$10$^{-13}$			& 3 \%			& no isobaric interference   	 \\
$^{237}$Np		 	& 2.14 Myr		& 		&  \textendash				& 5 \%			& no isobaric interference   	 \\
$^{239-244}$Pu	 	& kyr -- 80 Myr	& 		&  \textendash				& 2 \%			& no isobaric interference   	 \\
$^{241,243}$Am	 	& 430 \& 7370 yr	&  	&  \textendash				& 5 \%			& no isobaric interference  	 	 \\
$^{243-248}$Cm	 	& 18 yr -- 15.6 Myr	&  	&  \textendash				& 5 \%			& no isobaric interference   	 \\
superheavies	 		& 			& 			 	&  \textendash				& 				& no isobaric interference   	 \\
\hline
\end{tabular}\\
$^a$These are the best values achievable. The actual detection limits (given as isotope ratio radionuclide/stable isotope) and uncertainties may depend on the specific case and can be higher. The overall efficiency for the actinides is between 0.1\,\% and 1\,\%.
\end{table*}

Another ion may also mimic a true event in the detector. Contamination from chemistry or memory (cross contamination from other samples) in the ion source may induce an additional signal, misidentified as true event. Note, that impurities in the sample after chemical preparation are typically on the ppm-level (part per million, i.e. abundances at the $10^{-6}$ atom/atom level), whereas the rare isotope content is another 6 to 9 orders of magnitude lower. In order to quantify or check the background level, regular measurements of blank samples are crucial.

The overall efficiency (i.e. fraction of particles detected to that inserted into the ion source), which includes the efficiency for producing negative ions, stripping yield, transmission through the beam line and detector efficiency, depends strongly on the isotope under investigation. For some isotopes, e.g. carbon, chlorine or actinides, up to several percent can be obtained but in other cases one has to deal with an overall efficiency as low as 10$^{-3}$\,\%.

The beam intensity of the rare isotope is measured as count-rate with a particle detector, either multi-anode ionisation chambers, silicon strip detectors or surface barrier detectors, sometimes also coupled with time-of-flight detectors for suppression of isotopic interference. 
For AMS measurements, the isotope ratio is the relevant quantity with respect to signal to background and measurement accuracy. The conversion ratio (radionuclide/target nuclide, e.g.  $^{55}$Fe/$^{54}$Fe) is just a function of the cross section and the particle fluence (see eq.\,\ref{eq:sig}).

The sample itself is consumed in the subsequent AMS runs. The consumption rate is of the order of mg per hour under typical sputtering conditions, thus masses larger than 50 mg will not be useful, as measurement times much longer than days are not applicable. The situation becomes different for reaction products that are of a different element as the target nuclide (e.g. in case of charged particles in the entrance or exit channel, such as ($p,\gamma$) or ($\alpha,\gamma$) reactions). Because AMS usually measures isotope ratios of the same element (e.g. $^{55}$Fe relative to $^{54}$Fe and $^{56}$Fe), in this second case, following activation, the desired radionuclides will need to be separated from the bulk material, mixed (spiked) with milligrams of stable isotopes and converted into clean sputter samples for the subsequent AMS measurements. 

\begin{figure*}[ht]
\centering\includegraphics[angle=90,width=0.8\textwidth]{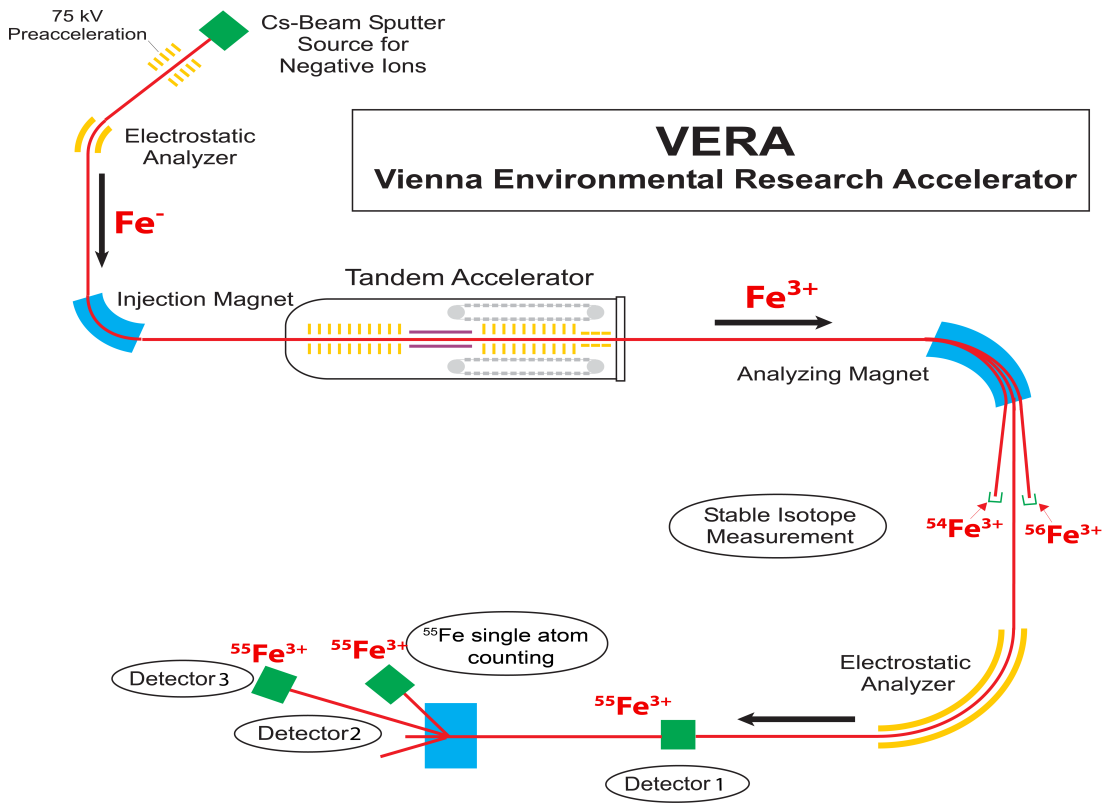}
\caption{Color online. Schematic layout of the AMS facility VERA. Negative Fe ions were extracted from the ion source and mass analyzed before the tandem accelerator.
After stripping in the terminal the 3-fold positively charged (3$^+$) ions  with an energy of 12 MeV were selected for analysis. The stable $^{54,56}$Fe nuclei were measured with Faraday cups, and the 
rare nuclide $^{55}$Fe was counted in one of the three subsequent particle detectors (see text for details).  
 \label{fig:vera}}
\end{figure*}

A schematic view of an AMS facility is shown in fig.\,\ref{fig:vera} including the detection devices for recording the stable isotopes and the low-intensity radionuclides. Negatively charged ions from a cesium sputter source are pre-accelerated and mass-analyzed in a low energy spectrometer. Isotopes of interest are sequentially injected as negative ions into the accelerator. By rapidly varying the respective particle energies of the different isotopes, the machine setup is adjusted for the different masses of the isotopes resulting in the same mass-energy product. In this way the particles can be adjusted to the same magnetic rigidity at the injection magnet (so-called beam sequencer, not shown in fig. \ref{fig:vera}) and consequently they follow sequentially the same beam trajectories.

In fig. \ref{fig:vera} the case of  $^{55}$Fe-AMS is shown, which is measured relative to the stable iron ions: the stable Fe isotopes are analyzed by current measurements with Faraday cups after the injection magnet and after the analyzing magnet (for $^{56}$Fe and $^{54}$Fe, respectively). The beam intensity of $^{55}$Fe is measured as count rate with one of the particle detectors. Such a sequence can be repeated 5 to 10 times per second with millisecond injection times for $^{54,56}$Fe, whereas the remaining 95\% of the time is used for $^{55}$Fe counting. The transmission through the accelerator is monitored by the currents measured at the low- and the high energy side. Because the measured $^{54}$Fe and $^{56}$Fe currents are defined by the isotopic composition of natural iron, the AMS runs of standards and irradiated samples can in this case be based on both, the $^{54}$Fe and 
the $^{56}$Fe beam. 

In a series of irradiations at KIT, neutron capture reactions for a 25 keV Maxwell-Boltzmann neutron energy distribution (see Tab. \ref{tab:xs_AMS_list}) were studied (see above), mainly relevant for s-process nucleosynthesis. Here a neutron flux density of $\sim$10$^9$ neutrons cm$^{-2}$ s$^{-1}$ was achieved on a typical target. After several days of activation, a neutron fluence of order (0.5 - 2)$\times$10$^{15}$ neutrons cm$^{-2}$ s$^{-1}$ could be accumulated. Combining this number with cross sections of order of 10 mbarn, we calculate a conversion ratio of $\sim$10$^{-11}$ (see Eq. \ref{eq:sig}). Such an isotope ratio is convenient for the routine AMS isotopes. 

As an example, we use the $^{54}$Fe($n, \gamma$)$^{55}$Fe cross section measurement again (see \cite{WBB17}), where the ratio of the $^{55}$Fe/$^{56}$Fe beam intensities, as produced in the irradiations, was of the order of 10$^{-11}$ to 10$^{-12}$. For comparison, the machine background of the $^{55}$Fe/$^{56}$Fe ratio was measured at the VERA facility to typically $<$$2\times10^{-15}$ atom/atom in the detector positions 2 and 3 (see fig. \ref{fig:vera}).

Detector position 1, located after the electrostatic analyzer, but before an additional magnetic filter, gave a $^{55}$Fe/$^{56}$Fe background of $2-3\times10^{-14}$. This higher background originated from a few $^{54}$Fe ions that were still accepted at this detector position. However, these ions were suppressed by more than a factor of 20 at the other detector positions further downstream. Overall, for this measurement the background contributed only less than 0.3 counts per hour to the observed $^{55}$Fe count rate of about one every few seconds. 

\begin{table*}[bt]
\caption{List of cross section studies that involved AMS measurements of the reaction products.
\label{tab:xs_AMS_list}}
\begin{tabular} {lrlllcl}
\hline\hline
Reaction				& 	AMS			& Irradiation    	& energy   			& AMS facility 		& meas. 	& reference	\\
Reaction				& 	isotope	 	&  facility   	&  range  				&  (terminal voltage)	&  uncertainty		& 	\\
\hline \\
$^{9}$Be($n, \gamma$)	& $^{10}$Be	& KIT	& 25 keV MB, 500 keV 		& VERA -- 3 MV			& 3 \%			& \cite{Wal10,WCD08}, in p. 	 \\
$^{13}$C($n, \gamma$)	& $^{14}$C	& KIT	& 25 keV MB, 120, 180 keV 	& VERA -- 3 MV			& 2--5 \%			& \cite{Wal10,WCD08},  	 \\
$^{14}$N($n, p$)		& $^{14}$C	& KIT	& 25 keV MB, 120, 180 keV 	& VERA -- 3 MV			& 2--5 \%			& \cite{Wal10,WCD08}, 	 \\
$^{26}$Mg($p, n$)		& $^{26}$Al	& ANL	& 5.2 -- 6.9 MeV		 	& ANL - tandem 			& 		& \cite{PHK80}			 \\
$^{25}$Mg($p, \gamma$)	& $^{26}$Al	& HZDR	& 189 - 408 keV			 	& TUM -- 14 MV			& 15 \%			&					\\
					&			&		&						& VERA -- 3 MV			& 3--10 \%		& \cite{AFF06}		 	 \\
$^{25}$Mg($p, \gamma$)	& $^{26}$Al	& LUNA	& 317 keV				 	& CIRCE -- 3MV			& 5 \%			& \cite{LSF10}			\\
$^{33}$S($\alpha, p$)	& $^{36}$Cl	& Notre Dame	& 0.7 -- 2.4 MeV/u	 	& Notre Dame -- 11MV	& 				& \cite{RGC12,ASC17,BKB13}			\\
$^{33}$S($\alpha, p$)	& $^{36}$Cl	& Notre Dame	& 0.8 -- 1.5 MeV/u 		& PRIME lab			& 5 \%			& \cite{RGC12,ASC17,BKB13}			\\
$^{36}$S($p, n$)		& $^{36}$Cl	& Notre Dame	& 			 		& Notre Dame -- 11MV	& 				& in p.			\\
$^{35}$Cl($n, \gamma$)	& $^{36}$Cl	& KIT	& 25 keV MB 				& VERA -- 3 MV			& 5 \%			& in p.  	 \\
$^{35}$Cl($n, \gamma$)	& $^{36}$Cl	& SARAF	& 35 keV MB 				& HZDR -- 6 MV			& 5 \%			& in p.  	 \\
					& 			&		&		 				& ANU -- 14 MV			& 5 \%			& in p.  	 \\	
$^{40}$Ca($n, \gamma$)	& $^{41}$Ca	& KIT	& 25 keV MB 				& VERA -- 3 MV			& 5 \%			&  \cite{DDH09} 	 \\
$^{40}$Ca($\alpha, \gamma$)	& $^{44}$Ti	& ATLAS	& $\sim$4.2 MeV resonances	& Weizmann -- 14 MV		& 			&  \cite{PFA03,NPA06,NPG05} 	 \\
$^{40}$Ca($\alpha, \gamma$)	& $^{44}$Ti	& Weizmann	& 2.1 -- 4.2 MeV integral	& Weizmann -- 14 MV		& 			&  \cite{NPA06,NPG05}, i. k. 	 \\
$^{40}$Ca($\alpha, \gamma$)	& $^{44}$Ti	& TUM		& 2.1 -- 4.17 integral and	& TUM -- 14 MV			& 7 \%			&  \cite{NBF06}, i. k. 	 \\
						& 			& 			& 4.17 -- 5.39 MeV integral \\
$^{40}$Ca($\alpha, \gamma$)	& $^{44}$Ti	& Notre Dame	& 3.5 -- 4.6 MeV 		& Notre Dame -- 11MV	& 		&  \cite{Bow13} 	 \\
$^{40}$Ca($\alpha, \gamma$)	& $^{44}$Ti	& HZDR		& MeV 			& HZDR -- 6 MV			&				&  in p. 	 \\
$^{52}$Cr($\alpha, n$)		& $^{55}$Fe	& Atomki		& 4.5 -- 10 MeV 	& ANU -- 14 MV			& 5 \%			&  in p.  	 \\
$^{54}$Fe($n, \gamma$)	& $^{55}$Fe	& KIT	& 25 keV MB, 450 keV			& VERA -- 3 MV			& 2--3 \%			&  \cite{WBB17} 	 \\
$^{58}$Ni($n, \gamma$)	& $^{59}$Ni	& KIT	& 25 keV MB 				& TUM -- 14 MV			& 8 \%			&  \cite{RDF07,LRD16} 	 \\
$^{62}$Ni($n, \gamma$)	& $^{63}$Ni	& KIT	& 25 keV MB 				& TUM -- 14 MV			& 10 \%			&  \cite{DFK10,RDF07} 	 \\
$^{64}$Ni($\gamma, n$)	& $^{63}$Ni	& HZDR	& 10.3 -- 13.5 MeV	 				& TUM -- 14 MV			& 10 \%			&  \cite{DFK10} 	 \\
$^{78}$Se($n, \gamma$)	& $^{79}$Se	& KIT	& 25 keV MB 				& TUM -- 14 MV			& 10 \%			&  \cite{RDF07} 	 \\
$^{92}$Zr($n, \gamma$)	& $^{93}$Zr	& SARAF	& 35 keV MB 				& ANU -- 14 MV			& 				& \cite{PCF18}, in p.  	 \\
$^{142}$Nd($\alpha, \gamma$)	& $^{146}$Sm	& Atomki		& 10 - 50 MeV 		& ANL -- Linac			& 				&  \cite{NBG16}, in p. 	 \\
$^{142}$Nd($\alpha, n$)	& $^{145}$Sm	& Atomki			& 10 - 50 MeV 		& ANL -- Linac			&				&  \cite{KPK12,K12,K08} 	 \\
$^{147}$Sm($\gamma, n$)	& $^{146}$Sm	& Tohoku U.	& MeV 			& ANL -- Linac			&				&  in p. 	 \\
$^{209}$Bi($n, \gamma$)	& $^{210}$Bi	& KIT	& 25 keV MB 				& VERA -- 3 MV			& 7 \%			&  in p.	 \\
$^{209}$Bi($n, \gamma$)	& $^{210}$Bi	& SARAF	& 35 keV MB 				&  ANL -- Linac			& 				&  in p.	 \\
\hline
\end{tabular}

Abbreviations: KIT, Karlsruhe Institute of Technology; VERA, the Vienna Environmental Research Accelerator, Univ. of Vienna; ANL, the Argonne National Laboratory; HZDR, the Helmholtz Center Dresden-Rossendorf; TUM, the Technical University Munich; LUNA, the Laboratory for Underground Nuclear Astrophysics at Gran Sasso; CIRCE, the Center for Isotopic Research on Cultural and Environmental heritage, Naples; Notre Dame, the Department of Physics, Univ. of Notre Dame, US; PRIME lab, the Purdue Rare Isotope Measurement Laboratory, US; SARAF, the Soreq Applied Research Accelerator Facility, Israel; ANU, the Heavy Ion Accelerator Laboratory (HIAF), the Australian National University; ATLAS, the the Argonne Tandem Linac Accelerator System (superconducting linear accelerator) at ANL; Weizmann, the Weizmann Institute of Science, Rehovot, Israel; Atomki, the Institute for Nuclear Research, Hungarian Academy of Sciences; Tohoku U., the Tohoku University, Japan; in p.: in progress; i.k.: inverse kinematics.
\end{table*}

\section{Summary}

Hopefully the readers of this review are convinced that activation is a powerful and versatile tool for cross section measurements in nuclear astrophysics. Essentially the only inevitable restriction of the activation method is the necessity of the residual nucleus of the reaction being radioactive. If this condition is met, different versions of activation can be applied to determine the cross section. The technique is practically suited for all kinds of astrophysically important reactions such as charged particle, neutron or $\gamma$-induced ones. Various methods are available also for the determination of the induced activity from the most commonly used $\gamma$-detection to several other ones. If long-lived isotopes are encountered, the extremely high sensitivity Accelerator Mass Spectrometry technique may be applied. 

For many decades the activation method has played a key role in experimental nuclear astrophysics. Many important reactions have been studied solely by this technique or as a complementary method to in-beam experiments. Clearly the importance of the method will not decrease in the future. There are still cases where the in-beam method cannot provide reliable cross section owing to e.g. high backgrounds (as in the case of the $\alpha$-capture reaction on heavy isotopes). In such cases, activation provides the only alternative to date. But even in cases where the in-beam technique is possible, the activation provides a useful independent approach in order to increase the precision and reliability of the results.

In the 21$^{\rm th}$ century, astronomical observations are entering an era of incredible precision. Experimental nuclear astrophysics must keep up with such a development in order to provide input data with the required precision to the astrophysical models used for the interpretation of the observations. Experimental techniques are also being continuously developed resulting in more powerful accelerators and more sophisticated detection instruments. Exploiting these developments the activation method can remain a valuable tool in the hands of nuclear astrophysics experimentalists.  

\section*{Acknowledgments}

The authors thank P. Mohr for the discussions on $\gamma$-induced reactions and for the comments on the manuscript, D. Balabanski for providing the information on the ELI-NP facility and A.J.T. Jull for the careful reading of the manuscript. This work was supported by NKFIH grants K120666 and NN128072 and by the \'UNKP-18-4-DE-449 New National Excellence Program of the Human Capacities of Hungary. G.G. Kiss acknowledges support form the J\'anos Bolyai research fellowship of the Hungarian Academy of Sciences.

%
%

\end{document}